\documentclass[12pt]{article}

\pdfoutput=1

\usepackage[top=80pt,bottom=85pt,left=67pt,right=67pt]{geometry}
\usepackage{amssymb,amsmath,xypic}
\usepackage{graphicx,color,subfigure}
\usepackage{float}
\usepackage{sectsty}
\usepackage{cite}
\usepackage[debug,pageanchor=false]{hyperref}
\definecolor{link}{rgb}{.8,.15,.1}
\hypersetup{colorlinks=true,linkcolor=link,citecolor=link,urlcolor=link,linktocpage}

\makeatletter
\@addtoreset{equation}{section}
\makeatother

\renewcommand{\theequation}{\thesection.\arabic{equation}}

\newcommand{\beq}{\begin{equation}}
\newcommand{\eeq}{\end{equation}}
\newcommand{\bea}{\begin{eqnarray}}
\newcommand{\eea}{\end{eqnarray}}
\newcommand{\nn}{\nonumber}

\begin{document}
\begin{titlepage}

\begin{center}

\vskip .5in 
\noindent

{\Large \bf{AdS$_3$ solutions in massive IIA, defect CFTs and T-duality}}

\bigskip\medskip

Yolanda Lozano$^{a,}$\footnote{ylozano@uniovi.es},  Niall T. Macpherson$^{b,c,}$\footnote{ntmacpher@gmail.com}, Carlos Nunez$^{d,}$\footnote{c.nunez@swansea.ac.uk}, Anayeli Ramirez$^{a,}$\footnote{anayelam@gmail.com} \\

\bigskip\medskip
{\small 

 $a$: Department of Physics, University of Oviedo,
Avda. Federico Garcia Lorca s/n, 33007 Oviedo, Spain
\vskip 3mm
 $b$: SISSA International School for Advanced Studies,
Via Bonomea 265, 34136 Trieste \\
		and \\INFN sezione di Trieste
		\vskip 3mm
$c$: International Institute of Physics, Universidade Federal do Rio Grande do Norte,
Campus Universitario - Lagoa Nova, Natal, RN, 59078-970, Brazil		
\vskip 3mm
 $d$: Department of Physics, Swansea University, Swansea SA2 8PP, United Kingdom}

\vskip .5cm 
\vskip .9cm 
     	{\bf Abstract }

\vskip .1in
\end{center}

\noindent
We establish a map between AdS$_3 \times$S$^2$ and AdS$_7$ solutions to massive IIA supergravity that allows one to interpret the former as holographic duals to D2-D4 defects inside 6d (1,0) CFTs.  This relation singles out in a particular manner the AdS$_3\times$S$^2$ solution constructed from AdS$_3\times$S$^3\times $CY$_2$ through non-Abelian T-duality, with respect to a freely acting SU(2). We find explicit global completions to this solution and provide
well-defined (0,4) 2d dual CFTs associated to them.  These
completions consist of linear
quivers with colour groups coming from D2 and D6 branes and flavour groups coming from D8 and D4 branes. Finally, we discuss the relation with flows interpolating between AdS$_3\times$S$^2\times$T$^4$ geometries and AdS$_7$ solutions found in the literature.
\noindent
 
\vfill
\eject

\end{titlepage}

\setcounter{footnote}{0}

\tableofcontents

\setcounter{footnote}{0}
\renewcommand{\theequation}{{\rm\thesection.\arabic{equation}}}

   \section{Introduction}

   Defect QFTs play an important role in our current understanding of Quantum Field Theories. Of particular interest is the situation when the ambient QFT is a CFT with a holographic dual. In this case, introducing appropriate branes in the dual geometry it is possible to construct the gravity dual of the defect QFT, that can then be studied holographically \cite{Karch:2001cw,Karch:2000gx,DeWolfe:2001pq}. When the defect QFT is a CFT, the explicit AdS dual geometry can be constructed in terms of the fully backreacted geometry \cite{DHoker:2007zhm,DHoker:2007hhe}, if the number of defect branes is sufficiently large.

  2d defect CFTs breaking half of the supersymmetries of the ambient CFT have been studied in \cite{Dibitetto:2017tve,Dibitetto:2017klx,Dibitetto:2018iar}, and their corresponding AdS$_3$ gravity duals have been constructed\footnote{1d CFTs and their AdS$_2$ duals have been addressed in \cite{Dibitetto:2018gtk}.}. The ambient CFT is either a 6d (1,0) CFT \cite{Dibitetto:2017tve,Dibitetto:2017klx} or a 5d fixed point theory \cite{Dibitetto:2018iar}\footnote{SUSY-preserving defects in 5d CFTs have been studied recently in \cite{Penin:2019jlf}.}. In the first case the 2d CFT lives in D2-D4 branes introduced in the D6-NS5-D8 brane intersections that underlie 6d (1,0) CFTs. In the second case it lives in D2-NS5-D6 branes in the D4-D8 brane set-ups that give rise to 5d Sp($N$) fixed point theories.
   
   In this work we will be interested in an extension of the first realisation. We will show that a sub-class of the local solutions constructed recently in \cite{LMNR1}, preserving small $\mathcal{N}=(0,4)$ supersymmetry on a foliation of  AdS$_3\times$S$^2\times$CY$_2$ over an interval, can be used to construct globally compact solutions dual to 2d (0,4) SCFTs that
   have an interpretation in terms of D2-D4 defects in 6d (1,0) CFTs. More precisely, we will be using the word defect to indicate the presence of extra branes in Hanany-Witten brane set-ups that would otherwise arise from compactifying higher dimensional branes. This provides a new scenario in which 2d (0,4) CFTs appear in string theory.
    
   2d (0,4) CFTs play a key role in the microscopical description of 5d black holes with AdS$_3\times$S$^2$ near horizon geometries  
\cite{Maldacena:1997de,Vafa:1997gr,Minasian:1999qn,Castro:2008ne,Haghighat:2015ega,Couzens:2019wls}. In string theory they can be realised in D1-D5-KK systems
\cite{Kutasov:1998zh,Sugawara:1999qp,Larsen:1999dh,Okuyama:2005gq} and D1-D5-D9 systems \cite{Douglas:1996uz}. They also play a prominent role in the description of self-dual strings in 6d (1,0) CFTs realised in M- and F-theory  \cite{Haghighat:2013gba,Haghighat:2013tka,Kim:2015gha,Gadde:2015tra,Lawrie:2016axq,Couzens:2017way}.
Their extensions to 2d (0,4) CFTs with large superconformal algebra have also received a good deal of attention \cite{Tong:2014yna,Lozano:2015bra,Kelekci:2016uqv,Hanany:2018hlz,Macpherson:2018mif}. Very recently we have also shown that they can be realised in larger D2-D4-D6-NS5-D8 brane systems \cite{LMNRPRL,LMNR2}.

In \cite{LMNR1} AdS$_3\times$S$^2\times$M$_4$ solutions in massive IIA supergravity preserving ${\mathcal N}=(0,4)$ supersymmetry with SU(2)-structure were classified. These solutions are warped products of AdS$_3\times$S$^2\times$M$_4$ over an interval, with M$_4$ either a CY$_2$ or a Kahler manifold. The CFT duals of the first class were studied in \cite{LMNRPRL,LMNR2}. They are described by (0,4) quiver gauge theories with gauge groups $\prod_{i=1}^n \text{SU}(k_i) \times \text{SU}({\tilde k}_i)$. $\text{SU}(k_i)$ is the gauge group associated to $k_i$ D2 branes stretched between NS5 branes and  SU(${\tilde k}_i$) is the gauge group associated to ${\tilde k}_i$ D6-branes, wrapped on the CY$_2$, also stretched between the NS5 branes. On top of these there are D4 and D8 branes that provide flavour groups to both types of nodes of the quiver. These quivers are a generalisation of the linear quivers studied in \cite{Gadde:2015tra}, where the D6 branes are unwrapped and are thus non-dynamical. In this paper we give an interpretation to our brane systems as D2-D4 brane defects in the D6-NS5-D8 branes associated to 6d (1,0) CFTs.

The organisation of the paper is as follows. In section 2 we review the main properties of the AdS$_3\times$S$^2\times$CY$_2$ solutions constructed in \cite{LMNR1}, and summarise the key features of their 2d dual CFTs, following \cite{LMNR2}. In section 3 we construct a mapping that relates a sub-class of these solutions with the AdS$_7$ solutions in massive IIA supergravity constructed in  \cite{Apruzzi:2013yva}. Using this map we can interpret the 2d dual CFTs as associated to D2-D4 defects  in the D6-NS5-D8 brane set-ups dual to the AdS$_7$ solutions, wrapped on the CY$_2$. This suggests that it should be possible to construct RG flows that interpolate between these two classes of solutions. In section 4 we discuss the AdS$_7$ solution that describes the 6d linear quiver with gauge groups of increasing ranks terminated by D6 branes, in relation to the map constructed in section 3. By means of this study we {\it rediscover} the non-Abelian T-dual (NATD) of the AdS$_3\times$S$^3\times$CY$_2$ geometry, constructed in \cite{Sfetsos:2010uq} (see also \cite{Lozano:2015bra}), as the {\it leading order} in an expansion on the number of gauge groups, of this solution. Then in section 5 we start a detailed study of the non-Abelian T-dual  solution. We show that it provides a simple explicit example in the general classification in \cite{LMNR1}, that describes a 2d (0,4) CFT with two families of gauge groups \cite{LMNR2} with increasing ranks.  As in other AdS solutions generated through non-Abelian T-duality, the solution is non-compact, and this renders and infinitely long dual quiver CFT. Remarkably, we are able to provide explicit global completions of the solution that have associated well-defined 2d (0,4) dual CFTs, that we describe. This solution thus provides a  useful example where it is possible to use holography in a very explicit way to determine global properties of non-compact solutions generated through non-Abelian T-duality, following the ideas in \cite{Lozano:2016kum,Lozano:2016wrs,Lozano:2017ole,Itsios:2017cew,Lozano:2018pcp}. In section 6 we attempt to make connection with RG flows in the literature that connect AdS$_3$ geometries in the IR, with an interpretation as 2d defect CFTs, with AdS$_7$ solutions in the UV 
 \cite{Dibitetto:2017tve,Dibitetto:2017klx}. Our results are negative, and thus exclude the RG flows constructed in these references as interpolating between the AdS$_3$ solutions in \cite{LMNR1} and the AdS$_7$ solutions in \cite{Apruzzi:2013yva}. Section 7 contains our conclusions and future directions. Appendix A contains some explicit derivations useful in section 5. Appendix B contains details of the BPS flow constructed in  \cite{Dibitetto:2017tve}, upon which section 6 is built.

 \section{AdS$_3\times$S$^2\times$CY$_2$ solutions in massive IIA and their CFT duals}
 
 In \cite{LMNR1} AdS$_3\times$S$^2$ solutions in massive IIA with small (0,4) supersymmetry and SU(2) structure were classified.  Two classes of solutions that are warped products of the form AdS$_3\times$S$^2\times$M$_4\times$I were found, for M$_4$ either a CY$_2$ manifold, class I, or a family of Kahler 4 manifolds depending on the interval, class II. The solutions in the first class provide a generalisation of D4-D8 systems involving additional branes, while those in the second class are a generalisation of the (T-duals of the) solutions in \cite{Couzens:2017way}, based on D3-branes wrapping curves in F-theory. In this paper we will be interested in the first class of solutions, that we now summarise. 
 
 The explicit form of the NS sector of the solutions referred as class I in \cite{LMNR1} is given by:
\begin{align}
ds^2&= \frac{u}{\sqrt{h_4 h_8}}\bigg(ds^2(\text{AdS}_3)+\frac{h_8h_4 }{4 h_8h_4+(u')^2}ds^2(\text{S}^2)\bigg)+ \sqrt{\frac{h_4}{h_8}}ds^2(\text{CY}_2)+ \frac{\sqrt{h_4 h_8}}{u} d\rho^2,\\
e^{-\Phi}&= \frac{h_8^{\frac{3}{4}} }{2h_4^{\frac{1}{4}}\sqrt{u}}\sqrt{4h_8 h_4+(u')^2},~~~~ H= \frac{1}{2}d(-\rho+\frac{ u u'}{4 h_4 h_8+ (u')^2})\wedge\text{vol}(\text{S}^2)+ \frac{1}{h_8}d\rho\wedge H_2.\nn
\end{align}
Here $\Phi$ is the dilaton, $H$ the NS 3-form and $ds^2$ is the metric in string frame. The warpings are determined from three independent functions $h_4,u,h_8$. $h_4$ has support on $(\rho,\text{CY}_2)$ while $u$ and $h_8$ have support on $\rho$, with $u'= \partial_{\rho}u$. The reason for the notation $h_4,h_8$ is that these functions may be identified with the warp factors of intersecting D4 and D8 branes  when $u=1$\footnote{The interpretation for generic $u$ is more subtle.}.\\ 
\\
The  10 dimensional RR fluxes are 
\begin{subequations}
\begin{align}
F_0&=h_8',\\[2mm]
F_2&=-H_2-\frac{1}{2}\bigg(h_8- \frac{ h'_8 u'u}{4 h_8 h_4+ (u')^2} \bigg)\text{vol}(\text{S}^2),\\[2mm]
F_4&= \bigg(d\left(\frac{u u'}{2 h_4}\right)+2 h_8  d\rho\bigg) \wedge\text{vol}(\text{AdS}_3)\nn\\[2mm]
& -\frac{h_8}{u} (\hat \star_4 d_4 h_4)\wedge d\rho- \partial_{\rho}h_4\text{vol}(\text{CY}_2)-\frac{u u'}{2 ( 4h_8 h_4+ (u')^2)} H_2\wedge \text{vol}(\text{S}^2),
\end{align}
\end{subequations}
with the higher fluxes related to these as $F_6=-\star_{10} F_4,~F_8=\star_{10} F_2,~F_{10}=-\star_{10} F_0$.\\
\\ 
Supersymmetry holds whenever
\beq\label{eq:caseIcon1}
u''=0,~~~~ H_2+ \hat{\star}_4 H_2=0,
\eeq
which makes $u$ a linear function. Here $\hat{\star}_4$ is the Hodge dual on CY$_2$. In turn, the Bianchi identities of the fluxes impose
\begin{align}\label{eq:BI1}
&h''_8= 0,~~~~dH_2=0\\[2mm]
&\frac{h_8}{u}\nabla^2_{\text{CY}_2}h_4+ \partial_{\rho}^2 h_4 -\frac{2}{h_8^3} \hat{\star}_4(H_2\wedge H_2)=0,
\nn
\end{align}
away from localised sources.

In this paper we will be interested in the subclass of solutions for which the symmetries of the CY$_2$ are respected by the full solution. This enforces $H_2=0$ and a compact CY$_2$. Thus,  we will be dealing with T$^4$ or K$3$. The supersymmetry and Bianchi identities are then all solved for $h_8, u, h_4$ arbitrary linear functions in $\rho$. 

The magnetic components of the Page fluxes $\hat{F}=F\wedge e^{-B_2}$, are given by 
\begin{eqnarray}
&&{\hat  f}_0=h_8', \label{fhat1} \\
&&{\hat f}_2=-\frac{1}{2}\Bigl(h_8-(\rho-2n\pi) h_8'\Bigr)\text{vol}(\text{S}^2) \\
&&{\hat f}_4=-h_4' \text{vol}(\text{CY}_2),\label{fhat4}\\
&& {\hat f}_6=\frac{1}{2}\Bigl(h_4-(\rho-2n\pi) h_4'\Bigr)\text{vol}(\text{CY}_2)\wedge \text{vol}(\text{S}^2) \label{fhat6},
    \end{eqnarray}
where we have included large gauge transformations of $B_2$ of parameter $n$, such that
\begin{equation}
B_2=\frac{1}{2}\Bigl(2n\pi-\rho+\frac{ u u'}{4 h_4 h_8+ (u')^2}\Bigr)\wedge\text{vol}(\text{S}^2).
\end{equation}

\begin{figure}
\centering
\includegraphics[scale=0.4]{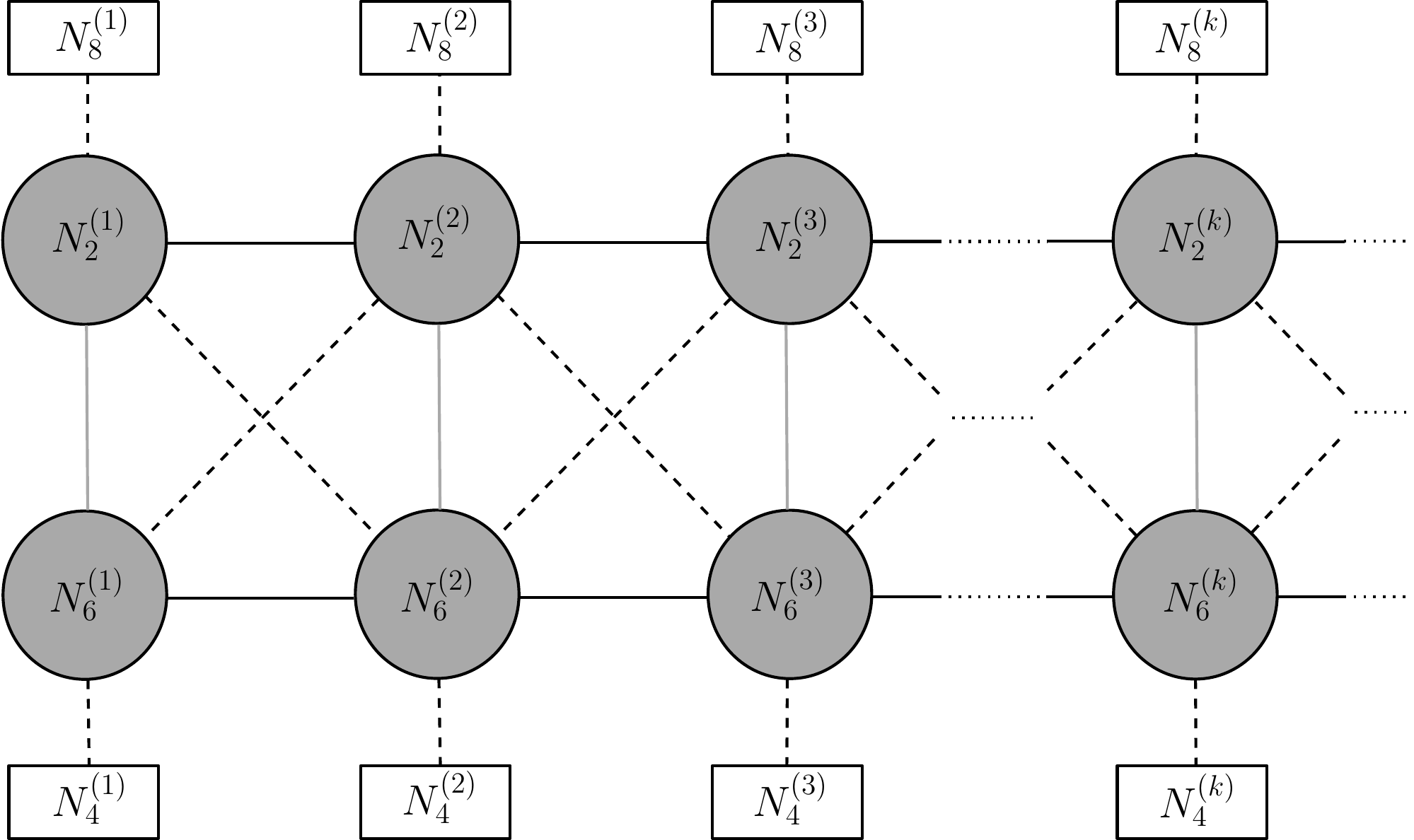}
\caption{Generic quiver field theory whose IR is holographic dual to the solutions discussed in this section. The solid black line represents a (4,4) hypermultiplet, the grey line a (0,4) hypermultiplet and the dashed line a (0,2) Fermi multiplet. (4,4) vector multiplets are the degrees of freedom at each node.}
\label{general-quiver}
\end{figure}

The 2d CFTs dual to this class of solutions were constructed in \cite{LMNR2}. They are described by 
 (0,4) supersymmetric quivers with gauge groups associated to D2 and D6 branes, the latter wrapped on the CY$_2$ manifold, stretched between NS5 branes. Having finite extension in this direction, the field theory living in both the D2 and D6 branes is two dimensional at low energies compared to the inverse separation between the NS5-branes. It was shown in \cite{LMNR2} that these quivers are rendered non-anomalous with adequate flavour groups at each node, coming from D4 and D8 branes. Remarkably, the flavour groups associated to gauge groups originating from D2 branes arise from D8 branes (wrapped on the CY$_2$) while those associated to the gauge groups originating from wrapped D6-branes arise from D4-branes. The corresponding quiver is depicted in figure \ref{general-quiver}. The underlying brane set-up is summarised in table \ref{D6-NS5-D8-D2-D4-first}. 

 The 2d CFTs dual to the solutions in class I thus generalise the (0,4) quivers studied in \cite{Gadde:2015tra} from D2, NS5 and D6 branes, in two ways. First, the D6 branes are compact, and therefore give rise to gauge, as opposed to global, symmetries. Second, there are D8 branes between the NS5 branes that can give rise to different flavour groups to each gauge group coming from D2 branes \cite{Hanany:1997gh,Brunner:1997gf}. Non-compact D4 branes provide the necessary flavour groups that render the nodes associated to the new, colour, D6 branes non-anomalous. Our quivers also generalise the (0,4) quivers constructed in  \cite{ Hanany:2018hlz} from D3-brane box configurations to gauge nodes with different gauge groups. 
     \begin{table}[ht]
	\begin{center}
		\begin{tabular}{| l | c | c | c | c| c | c| c | c| c | c |}
			\hline		    
			& 0 & 1 & 2 & 3 & 4 & 5 & 6 & 7 & 8 & 9 \\ \hline
			D2 & x & x & &  &  &  & x  &   &   &   \\ \hline
			D4 & x & x &  &  &  &   &  & x & x & x  \\ \hline
			D6 & x & x & x & x & x & x & x  &   &   &   \\ \hline
			D8 & x & x &x  & x & x &  x &  & x & x & x  \\ \hline
			NS5 & x & x &x  & x & x & x  &   &   &  &  \\ \hline
		\end{tabular} 
	\end{center}
	\caption{$\frac18$-BPS brane intersection underlying the quiver depicted in figure \ref{general-quiver}. $(x^0,x^1)$ are the directions where the 2d CFT lives, $(x^2, \dots, x^5)$ span the CY$_2$, on which the D6 and the D8-branes are wrapped, $x^6$ is the direction along the linear quiver, and $(x^7,x^8,x^9)$ are the transverse directions on which the $SO(3)_R$ symmetry is realised.}   
	\label{D6-NS5-D8-D2-D4-first}	
\end{table} 


$\frac18$-BPS brane set-ups such as the one depicted in table \ref{D6-NS5-D8-D2-D4-first} were discussed in \cite{Dibitetto:2017klx} in the context of 2d defect CFTs originating from D2-D4 branes living in 6d (1,0) CFTs. In the next section we find that it is indeed possible to give an interpretation to some of the CFTs studied in \cite{LMNR2} in these terms. We will discuss the connection with the solutions constructed in \cite{Dibitetto:2017klx} in section 6.

\section{A map between AdS$_3\times$S$^2$ and AdS$_7$ solutions in massive IIA}   
   \label{AdS7limit}
   
 In \cite{Apruzzi:2013yva} an infinite class of AdS$_7$ solutions in massive IIA was constructed\footnote{See \cite{Apruzzi:2017nck} for orientifold constructions thereof.}, preserving 16 supersymmetries (eight Poincare and eight conformal) on a foliation of AdS$_7\times$S$^2$ over an interval. 
 In this section we show that they can be related to our solutions in \cite{LMNR1}, preserving (0,4) supersymmetries on a foliation of AdS$_3\times$S$^2\times$CY$_2$ over an interval, through a map that reduces supersymmetry by half. As opposed to the mappings in \cite{Apruzzi:2015wna} between  AdS$_7$ solutions and the AdS$_5$ and AdS$_4$ solutions in \cite{Apruzzi:2015zna,Rota:2015aoa}, this mapping is not one-to-one, due to the presence of D2-D4 defects, whose backreaction introduces new 4-form and 6-form fluxes.

 We start by briefly summarising the solutions constructed in  \cite{Apruzzi:2013yva}. 
 Using the parametrisation in \cite{Cremonesi:2015bld}, these solutions can be completely determined by a function $\alpha(z)$ that satisfies the differential equation
  \begin{equation}
\label{dddotalfa}
\dddot{\alpha}=-162\pi^3 F_0.
\end{equation}
Where $F_0$ is the Ramond zero-form. Explicitly, the metric and fluxes are given by
\begin{eqnarray}
\label{metricAdS7alpha}
ds_{10}^2&=&\pi\sqrt{2} \bigg(8\sqrt{-\frac{\alpha}{\ddot{\alpha}}} ds^2(\text{AdS}_7)+\sqrt{-\frac{\ddot{\alpha}}{\alpha}}dz^2+\frac{\alpha^{3/2}(-\ddot{\alpha})^{1/2}}{\dot{\alpha}^2-2\alpha\ddot{\alpha}}ds^2(\text{S}^2)\bigg) \\
e^{2\Phi}&=&2^{5/2}\pi^5 3^8\frac{(-\alpha/\ddot{\alpha})^{3/2}}{\dot{\alpha}^2-2\alpha\ddot{\alpha}}\\
B_2&=&\pi \Bigl(-z+\frac{\alpha\dot{\alpha}}{\dot{\alpha}^2-2\alpha\ddot{\alpha}}\Bigr)\text{vol}(\text{S}^2) \\
\label{F2AdS7alpha}
F_2&=&\Bigl(\frac{\ddot{\alpha}}{162\pi^2}+\frac{\pi F_0\alpha\dot{\alpha}}{\dot{\alpha}^2-2\alpha\ddot{\alpha}}\Bigr)\text{vol}(\text{S}^2).
\end{eqnarray}
These backgrounds were shown to arise as near horizon geometries of D6-NS5-D8 brane intersections \cite{Bobev:2016phc,Macpherson:2016xwk} (see also \cite{Cremonesi:2015bld,Gaiotto:2014lca} for previous hints), from which 6d linear quivers with 8 supercharges can be constructed \cite{Hanany:1997gh,Brunner:1997gf}.  In these quivers anomaly cancelation implies that for every gauge group the number of flavours must double the number of gauge multiplets, $N_f=2N_c$ \cite{Gaiotto:2014lca}. In reference \cite{Cremonesi:2015bld} a prescription was given to calculate the function $\alpha(z)$ that encodes the explicit AdS$_7$ solution dual to a given 6d quiver diagram. In this quiver diagram the NS5 branes are located at different values of $z$, the D6-branes are stretched between them along this direction and the D8 branes are perpendicular. The corresponding brane set-up is depicted in table \ref{D6-NS5-D8}.
 \begin{table}[ht]
	\begin{center}
		\begin{tabular}{| l | c | c | c | c| c | c| c | c| c | c |}
			\hline		    
			& 0 & 1 & 2 & 3 & 4 & 5 & 6 & 7 & 8 & 9 \\ \hline
			D6 & x & x & x & x & x & x & x  &   &   &   \\ \hline
			D8 & x & x &x  & x & x &  x &  & x & x & x  \\ \hline
			NS5 & x & x &x  & x & x & x  &   &   &  &  \\ \hline
		\end{tabular} 
	\end{center}
	\caption{$\frac14$-BPS brane intersection underlying the 6d (1,0) CFTs dual to the AdS$_7$ solutions in \cite{Apruzzi:2013yva}. $(x^0,\dots,x^5)$ are the directions where the 6d CFT lives, $x^6$ is the direction along which the NS5-branes are located, and $(x^7,x^8,x^9)$ realise the SU(2) R-symmetry of the internal space.}   
	\label{D6-NS5-D8}	
\end{table} 

After this brief summary we can introduce the mapping that relates these solutions to the solutions in class I in \cite{LMNR1}, summarised in the previous section. The mapping reads
  \begin{eqnarray}
  \label{fieldthdir}
  \rho&\leftrightarrow&2\pi z\\
  \label{ucuadrado}
    u&\leftrightarrow&\alpha\\
    \label{h8alphaddot}
    h_8&\leftrightarrow&-\frac{\ddot{\alpha}}{81 \pi^2} \\
    \label{ultima}
    h_4&\leftrightarrow& \frac{81}{8} \alpha\, .
    \end{eqnarray}
   Using these relations one can match the $B_2$ field, dilaton, $F_0$ and $F_2$ fluxes of the two solutions, as well as  the $S^2\times I$ components of the metric. For the rest of the metric one must consider the mapping
    \begin{equation}
    ds^2(\text{AdS}_3)+\frac{3^4}{2^3}ds^2(\text{CY}_2)\leftrightarrow \,4\, ds^2(\text{AdS}_7)\,.
    \end{equation}
 Besides, the $F_4$ and $F_6$ fluxes,  which would violate the symmetries of the AdS$_7$ solution, must be disregarded when using the mapping from AdS$_3$ to AdS$_7$. These fluxes clearly sign the presence of a D2-D4 defect in the AdS$_3$ solution. As we discuss below, its backreaction has also the effect of modifying the dependence of the different functions on both sides of (\ref{ucuadrado})-(\ref{ultima}) on the respective {\it field theory} directions (related through (\ref{fieldthdir})).

 Indeed, (\ref{ucuadrado}) and (\ref{ultima}) relate linear functions in $\rho$ with a cubic function of $z$. This mapping is therefore essentially different from the mappings found in 
\cite{Apruzzi:2015wna}, where other than the replacements of AdS$_5\times\Sigma_2$ or AdS$_4\times \Sigma_3$ with AdS$_7$, the internal space is just distorted by some numerical factors. This difference is due to  the presence of the D2-D4 defect in the AdS$_3$ solution, which is also responsible for the reduction of the
supersymmetry from 1/2 BPS to 1/4  BPS. 

Using (\ref{h8alphaddot}) and (\ref{fieldthdir}) it is possible to obtain
the AdS$_7$ solution related to a particular AdS$_3\times$CY$_2$ solution. One finds
  \begin{equation}
    \label{h8ddotalpha}
    h_8=F_0\rho+c\quad \leftrightarrow \quad \ddot{\alpha}=-162\pi^3 F_0 z + {\tilde c}\, ,
    \end{equation}
    from which $\alpha(z)$, and thus, the explicit AdS$_7$ solution in  \cite{Apruzzi:2013yva}, can be determined. This mapping does not however give the expressions for the $u$ and $h_4$ functions that define the AdS$_3$ solution. Still, one can exploit (\ref{h8ddotalpha}) to show that  the D8-brane charges of the AdS$_7$ and AdS$_3$ solutions, determined, respectively, from  $h_8'$ and $-\dddot{\alpha}/(162\pi^3)$, agree, and that the same holds for the D6-brane charges, given that the corresponding ${\hat f}_2$ Page fluxes satisfy
         \begin{equation}
    {\hat f}_{2(\text{AdS}_3)}=-\frac{1}{2}\Bigl(h_8-(\rho-2n\pi) h_8'\Bigr)\text{vol}(S^2)\leftrightarrow\Bigl(\frac{\ddot{\alpha}}{162 \pi^2}+F_0 (z-n\pi)\Bigr)\text{vol}(S^2)= {\hat f}_{2(\text{AdS}_7)}.
    \end{equation}
    This implies  that the  D6-NS5-D8 {\it sector} of the AdS$_3$ solution is simply
obtained by compactifying on the CY$_2$ the D6-NS5-D8 branes that underlie the AdS$_7$ solution.
     
     However, as we have mentioned, the $u$ and $h_4$ linear functions needed to fully specify the AdS$_3$ solution, cannot be determined from the AdS$_7$ solution using this mapping, other than the fact that they have to be proportional to each other\footnote{We will see below that this guarantees that the two solutions share the same singularity structure, or, in other words, that the S$^2$ shrinks in the same way to produce topologically an S$^3$.}. This was to be expected, since, as we showed in \cite{LMNR1}, these functions encode the information of the additional D2-D4 branes present in the AdS$_3$ solution.   This is, once more, essentially different from the mappings between AdS$_5$ and AdS$_4$ and AdS$_7$ solutions found in \cite{Apruzzi:2015wna}, where it is not possible to identify 6 and 4-cycles on which additional D2 or D4 brane charges can be defined. In this case this is possible due to the non-trivial CY$_2$ 4-cycle in the internal space of the AdS$_3$ solutions.
    
      The symmetry between the D6-NS5-D8 and D2-NS5-D4 {\it sectors}, manifest in the expressions of the RR Page fluxes of the AdS$_3$ solutions,
    \begin{equation}
    \hat{f}_0=h_8',\qquad {\hat f}_2=-\frac{1}{2}\Bigl(h_8-(\rho-2n\pi) h_8'\Bigr)\text{vol}(\text{S}^2)
    \end{equation}
    and 
    \begin{equation}
    \label{F4F6}
    {\hat f}_4=-h_4' \text{vol}(\text{CY}_2),\qquad {\hat f}_6=\frac{1}{2}\Bigl(h_4-(\rho-2n\pi) h_4'\Bigr)\text{vol}(\text{CY}_2)\wedge \text{vol}(\text{S}^2),
    \end{equation}
stress the role of both
 D2 and D6 branes as colour branes in the 2d CFT dual to the AdS$_3$ solution, and of D4 and D8 branes as flavour branes \cite{LMNR2}. The resulting 2d (0,4) CFT thus contains two types of nodes, associated to the gauge groups of D2 and compact, wrapped on the CY$_2$, D6 branes. This is the generalisation of the (0,4) quivers discussed in \cite{Gadde:2015tra} that we found in \cite{LMNR2}. Note that  compactification on the CY$_2$ of the 6d CFT living in D6-NS5-D8 branes preserves 
(4,4) supersymmetries\footnote{Gauge theories with (4, 4) supersymmetry in two dimensions may be viewed as the dimensional reduction of 6d (1, 0) gauge theories. The six dimensional gauge theories have an $SU(2)_R$ R-symmetry. Upon dimensional reduction to two dimensions there is an additional $\text{SO}(4)=\text{SU}(2)_r \times \text{SU}(2)_l$ symmetry acting on the four reduced dimensions. This is also an R-symmetry since the supercharges are a spinor of this SO(4) group; the left-moving (positive chirality) supercharges are in the (2,1,2) representation of $\text{SU}(2)_l \times \text{SU}(2)_r \times \text{SU}(2)_R$ while the right-moving (negative chirality) supercharges are in the (1, 2, 2) representation \cite{Witten:1997yu,Diaconescu:1997gu}.}. The D2-D4 branes further reduce the supersymmetries by one half  \cite{Dibitetto:2017klx} (see also  \cite{Boonstra:1998yu}).
Alternatively, one could start with the D2-NS5-D4 Hanany-Witten brane set-ups discussed in \cite{Brodie:1997wn,Ito:1997ep}, realising 2d (4,4) field theories, and intersect them with wrapped D6 and D8 branes, which would also reduce the supersymmetries by a half.  The resulting $\frac{1}{8}$ BPS configuration (increasing to $\frac{1}{4}$ at the near horizon) is the one that we depicted in table \ref{D6-NS5-D8-D2-D4-first}.

Let us now discuss the physical reason for the condition $h_4\sim u$, implied by   (\ref{ultima}) and (\ref{ucuadrado}), in the AdS$_3$ solutions.  As we have mentioned, the functions $u$ and $h_4$, needed to completely determine the AdS$_3\times$CY$_2$ solution, cannot be computed  from (\ref{ucuadrado}) and (\ref{ultima}), due to the different dependence on $\rho$ and $z$ of these functions and $\alpha(z)$, respectively. Rather, the relation  $h_4=81 u/8$ has to be seen as a restriction on the class of AdS$_3\times$CY$_2$ solutions that can be interpreted as defects in the CFTs dual to AdS$_7$ solutions. This restriction comes from the condition that
 both solutions share the same singularity structure. In order to see this we note that
in both solutions the range of the interval is determined by the points at which the S$^2$ shrinks, such that the S$^2\times$I space is topologically an S$^3$. In AdS$_7$ there is a D6 brane when $\alpha=0$, $\ddot{\alpha}\neq 0$, and a O6 when $\ddot{\alpha}=0$, $\alpha\neq 0$. In turn, when 
 $\alpha=0$, $\ddot{\alpha}= 0$ the S$^2$ shrinks smoothly \cite{Apruzzi:2015zna}.
  Similarly, for AdS$_3$ solutions satisfying $h_4\sim u$  there is a D6 brane when $u\sim h_4=0$, $h_8\neq 0$ and a O6 when $h_8=0$, $u\sim h_4\neq 0$. In turn, the $S^2$ shrinks smoothly for $u\sim h_4=0$, $h_8=0$ \cite{LMNR1}. The role played by the D6 branes terminating the space as flavour branes is discussed in section \ref{linear-quiver}.
	
Let us summarise our findings so far in this section.  We have shown that a subclass of the solutions in \cite{LMNR1}\footnote{Those that share the same singularity structure of the solutions in \cite{Apruzzi:2013yva}, in the sense that we have just explained.} can be interpreted as arising from D2-D4 {\it defect branes} inside the D6-NS5-D8 brane intersections underlying the AdS$_7\times \text{S}^2\times I$ solutions in \cite{Apruzzi:2013yva}, wrapped on the  CY$_2$ of the internal manifold. 6d (1,0) CFTs compactified in CY$_2$ manifolds give rise to 2d (4,4) field theories that are not conformal \cite{Witten:1997yu,Diaconescu:1997gu}. Therefore, AdS$_3$ solutions cannot be obtained from the AdS$_7$ solutions in \cite{Apruzzi:2013yva} simply by extending the construction of AdS$_5$ and AdS$_4$ solutions in \cite{Apruzzi:2015wna}  to 4d manifolds. As we showed in \cite{LMNR1} extra D2 and D4 branes are needed, that further reduce the supersymmetries down to 1/8 BPS and the AdS$_3$ solutions to 1/4-BPS. These branes backreact in the compactified geometry, and modify the simple mappings found in \cite{Apruzzi:2015wna} such that the dependence of the functions defining the AdS$_3$ and AdS$_7$ solutions change, due to the backreaction. One can thus think of the 2d CFT associated to the AdS$_3$ solutions as comprised of two sectors, one coming from D6-NS5-D8 branes wrapped on the CY$_2$, which by itself does not give rise to a 2d CFT, and one coming from extra, D2-D4 branes, which would not give rise either to 2d CFTs together with the NS5-branes \cite{Brodie:1997wn}. One can in this sense interpret the D2-D4 branes as defects inside D6-NS5-D8 brane systems. We would like to stress that this defect interpretation is essentially different from the defect interpretation in terms of punctures that can be given to the Gaiotto theories in 4d \cite{Gaiotto:2009we}, dual to the Gaiotto-Maldacena geometries \cite{Gaiotto:2009gz}. In this last case both the field theory in the absence of punctures (dual to the Maldacena-Nunez solution \cite{Maldacena:2000mw}) and the ones with punctures are well- defined 4d CFTs, in contrast with the 2d CFTs dual to our AdS$_3$ solutions.
  
Further light on the relation between the 2d (0,4) CFTs dual to the AdS$_3$ solutions and compactifications on CY$_2$ of the 6d (1,0) CFTs dual to the AdS$_7$ solutions comes from comparing their respective central charges, following  \cite{Bobev:2017uzs}.
  The holographic central charge of the 6d CFTs dual to the AdS$_7$ solutions was computed in  \cite{Nunez:2018ags}:
  \begin{equation}
 c_{\text{AdS}_7}=\frac{1}{G_N}\frac{2^4}{3^8}\int dz (-\alpha \ddot{\alpha}).
 \end{equation}
 In turn, the holographic central charge of the 2d CFTs dual to the AdS$_3\times$CY$_2$ solutions is  \cite{LMNR2}
 \begin{equation}
 \label{cads3}
 c_{\text{AdS}_3}=\frac{3\pi}{2G_N}{\rm Vol}(\text{CY}_2)\int d\rho (h_8h_4)\, .
 \end{equation}
Using the mapping given by (\ref{fieldthdir})-(\ref{ultima}) this becomes
 \begin{equation}
 \label{ces}
 c_{\text{AdS}_3}\leftrightarrow\frac{3}{2^3 G_N}{\rm Vol}(\text{CY}_2)\int dz (-\alpha \ddot{\alpha})=\frac{3^9}{2^7}
 {\rm Vol}(\text{CY}_2) \,c_{\text{AdS}_7}\, .
 \end{equation}
 Thus, there exists a universal relation between the central charges associated to both types of solutions. Similarly, 
   in \cite{Passias:2015gya} (see also \cite{Benini:2013cda}) AdS$_3\times \Sigma_4$ solutions of massive IIA were constructed whose 2d
 (0,1) and (0,2) CFT duals arise as compactifications of the 6d (1,0) theories dual to the AdS$_7$ solutions. Their respective free energies were shown to satisfy the relation  
    \begin{equation}
  \frac{{\cal F}_2}{{\cal F}_6}=\frac{1}{(2X_{IR})^5}{\rm Vol}(\Sigma_4),
  \end{equation}
where $\Sigma_4$ is the compactification manifold and $X_{IR}$ is a constant that characterises the AdS$_3$ solution\footnote{$X_{IR}$ is the value in the IR of the $X$ scalar field of 7d minimal supergravity (see section 6).}.      
     Our result is thus in agreement with an interpretation of the 2d CFTs dual to our solutions as compactified  6d (1,0) theories in CY$_2$ manifolds, with extra degrees of freedom coming from the 2d  defects. It  would be very interesting to obtain explicit flows connecting the AdS$_3\times$CY$_2$ solutions in the IR with the AdS$_7$ solutions in the UV. In particular, it would be interesting to clarify whether these involve $\mathbb{R}_{1,1}\times$CY$_2$ warped product geometries, which would be the natural extension of the flows constructed in \cite{Benini:2013cda,Passias:2015gya,Bobev:2017uzs}, or wrapped AdS$_3$ subspaces, more directly related to defects, as in 
  \cite{Dibitetto:2017tve,Dibitetto:2017klx,Dibitetto:2018iar}.     In \cite{Dibitetto:2017klx}
different limits of the D2-D4-D6-NS5-D8 intersections depicted in table \ref{D6-NS5-D8-D2-D4-first} were studied, giving rise to either AdS$_7$ or AdS$_3 \times$S$^3 \times$I' geometries, associated to the UV or IR limits of the intersection, respectively. In particular, AdS$_3\times$T$^4$ geometries should arise when the branes are smeared on the T$^4$. In section 6 we explore the connection between the 
  BPS flows constructed in 
  \cite{Dibitetto:2017tve,Dibitetto:2017klx} and the subclass of AdS$_3\times$T$^4$ solutions defined by the mapping discussed in this section.


\section{The linear quiver with infinite number of nodes}\label{linear-quiver}
  
As we have mentioned, the mapping found in the previous section is formal, in the sense that it relates $\alpha$, a cubic function in $z$, to $h_4\sim u$, which are linear in $\rho$ (with $z$ and $\rho$ related as in (\ref{fieldthdir})).  In this section we discuss a particular instance in which $\alpha$ and $h_4\sim u$ can be explicitly related.

\begin{figure}
\centering
\includegraphics[scale=0.9]{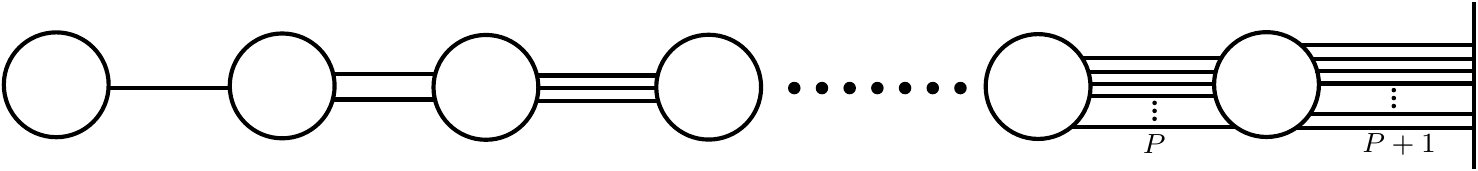}
\caption{D6-NS5-D8 brane set-up associated to a linear quiver with increasing ranks terminated by a flavour group. NS5 branes are denoted by circles, D6 branes by horizontal lines and D8 branes by vertical lines.}
\label{Branesetup}
\end{figure}

Consider an AdS$_7$ solution in which the S$^2\times$I geometry is smooth at $z=0$ and terminates at $z=P+1$, such that
 \begin{equation} \label{baticani}
F_0= -\frac{\alpha'''(z)}{162 \pi^3}
                    = \frac{ N}{2\pi}  \left\{ \begin{array}{ccrcl}
                     1\,,
                       & 0\leq z\leq P \\
                        -P\,, &~~ P\leq z \leq P+1.
                                             \end{array}
\right.
\end{equation}
For this we need $N(P+1)$ D8-branes at $z=P$, given that
\begin{equation}
dF_0= \frac{N(P+1)}{2\pi}\delta(z-P) dz.
\end{equation}
As shown in \cite{Nunez:2018ags}, for a particular choice of the integration constants such that $\alpha(0)=\alpha(P+1)=0$, and $\alpha$ and $\alpha'$ are continuous functions, we have
  \begin{equation} \label{bati}
\alpha(z)
                    =\frac{27\pi^2 N}{2}\left\{ \begin{array}{ccrcl}
                      P(P+2)z  -z^3\,,
                       & 0\leq z\leq P \\
                        Pz^3-3P(P+1)z^2+P(3P^2+4P+2)z-P^3(P+1)\, , &~~ P\leq z \leq P+1,
                                             \end{array}
\right.
\end{equation}
and the dual CFT is a linear quiver with gauge group
\begin{equation}
\text{SU}(N)\times \text{SU}(2N)\times \text{SU}(3N)\times \text{SU}(4N)\times...\times \text{SU}(PN),\label{beto}
\end{equation}
finished with a $\text{SU}((P+1)N)$ flavour group, represented by the D8 branes.  The brane set-up associated to this quiver is depicted in figure \ref{Branesetup}.

Now, consider the situation in which $P$ is very large, so that the region of interest reduces to $0\leq z\leq P$ and we can take $\alpha(z)=\frac{27\pi^2 N}{2} (P(P+2)z  -z^3)$ for all $P$ \footnote{Note that strictly speaking this would extend the region of interest to $0\leq z \leq \sqrt{P(P+2)}$, but this is equivalent to $0\leq z\leq P$ when $P$ is large.}.
Redefining $z= \sqrt{P(P+2)} x$, we can write the solution in this region as
\begin{eqnarray} \label{AdS7x}
& &\frac{ds^2}{\sqrt{P(P+2)}}= \frac{8\pi}{\sqrt{3}}\sqrt{1-x^2}\, ds^2(\text{AdS}_7) + \frac{2\sqrt{3}\pi}{\sqrt{1-x^2}}\Big[dx^2+ \frac{x^2(1-x^2)^2}{1+6x^2-3x^4}\, ds^2(\text{S}^2) \Big],\nonumber\\
& & e^{4\Phi} =\frac{12}{F_0^4\pi^2 P(P+2)} \frac{(1-x^2)^3}{(1+6x^2-3x^4)^2},\;\;\; F_0=\frac{N}{2\pi}\\
& & B_2=-2\pi \sqrt{(P(P+2)}  \frac{x^3(5-3x^2)}{1+6x^2-3x^4}\, {\rm vol}(\text{S}^2),\;\;\; F_2= F_0 B_2,\;\;\; \hat{f}_2= n {\rm vol}(\text{S}^2).\nonumber
\end{eqnarray}
This solution can be expanded close to $x=1$ (the end of the space) by defining $x=1-v$. We then have a metric and dilaton that for small values of $v$ read,
\begin{eqnarray}
& & ds^2\sim 8\pi\sqrt{\frac{2}{3}} \sqrt{v}\, ds^2(\text{AdS}_7) + \frac{\sqrt{6} \pi}{\sqrt{v}} (dv^2+ v^2 ds^2(\text{S}^2)),\nonumber\\
& & e^{4\Phi}\sim v^3.
\end{eqnarray}
It is thus clear that close to $v\sim 0$ or $x\sim 1$, in the end of the space, we have D6 branes that extend along AdS$_7$. As discussed in \cite{Apruzzi:2015zna}, these D6 branes can play the role of flavour branes even when their dimensionality is the same as that of the colour branes. They differ in that the colour branes are extended along the six Minkowski directions of AdS$_7$ plus a bounded interval, while the flavour D6-branes are extended on the whole AdS$_7$. Being non-compact they can act as flavour branes, as  happens in many other (qualitatively different) examples, like \cite{Casero:2006pt}.

Now, we would like to use the mapping between AdS$_3$ and AdS$_7$ solutions described by (\ref{fieldthdir})-(\ref{ultima}). This tells us that we should identify,
\begin{eqnarray}
h_8= \frac{N}{2\pi} \rho,\;\;\;\; u=  \frac{27\pi N}{4} (P(P+2)\rho -\frac{\rho^3}{4\pi^2}),\;\;\;\; h_4=\frac{81}{8} u.
\end{eqnarray}
This is not a solution of the equations of motion of the AdS$_3$ system. Nevertheless, if we take $P\to\infty$ or, equivalently, $\rho\to 0$, we have
\begin{equation}
h_8=\frac{N}{2\pi} \rho,\;\;\; u= \frac{27\pi N}{4}P(P+2) \rho,\;\;\;\; h_4=\frac{3^7 \pi N}{2^5}P(P+2) \rho,\label{carlitos}
\end{equation}
which defines a non-compact AdS$_3$ solution. This is 
 the solution constructed in \cite{Sfetsos:2010uq} acting with non-Abelian T-duality on the AdS$_3\times$S$^3\times$CY$_2$ solution dual to the D1-D5 system \cite{Witten:1997yu,Diaconescu:1997gu,Giveon:1998ns,Seiberg:1999xz,Aharony:1999dw}. 
 
 As we discuss in the next section,
 the non-compact nature of the non-Abelian T-dual solution is reflected in the dual CFT in the existence of an infinite number of gauge groups of increasing ranks. In this section we have {\it rediscovered} it as the {\it leading order} of the solution defined by \eqref{AdS7x}, dual to a well-defined six dimensional CFT\footnote{To be more precise, (\ref{carlitos}) selects a particular non-Abelian T-dual solution, with a given relation between the D2 and D6 brane charges. We give more details in the next section.}.
Since we are working at very small values of $z$ (equivalently, very small values of $\rho$), we do not see the flavour D6 branes, and the space is rendered non-compact. Conversely, taking $P\to\infty$ we see no sign of these branes closing the space.

We discuss the non-Abelian T-dual solution in detail in the next section, and describe other possible ways to define it globally using AdS$_3$/CFT$_2$ holography.



   \section{The non-Abelian T-dual of AdS$_3\times$S$^3\times$CY$_2$} \label{NATD}
   
   In this section we discuss in detail one of the simplest solutions in the classification of  AdS$_3\times$S$^2$ geometries in \cite{LMNR1}, with a focus on the description of its 2d dual CFT, following \cite{LMNR2}. This solution arises  acting with non-Abelian T-duality on the near horizon of the D1-D5 system, and was originally constructed in \cite{Sfetsos:2010uq}. In reference \cite{Lozano:2015bra} it was shown that the (4,4) supersymmetry of the D1-D5 system is reduced to (0,4) upon dualisation, and that the solution can be further T-dualised and uplifted to M-theory such that it fits in the class of AdS$_3\times$S$^2\times$S$^2\times$CY$_2$ solutions in \cite{Kim:2007hv}\footnote{Actually, it provides the only known example in this class with SU(2) structure.}. This solution is particularly interesting in the study of the interplay between non-Abelian T-duality and holography, since it allows for simple explicit global completions of the geometry using field theory arguments.
   
 In this section we also discuss another solution in the class in \cite{LMNR1} that arises from the D1-D5 system, and that can be obtained as a limit of the non-Abelian T-dual solution \cite{Macpherson:2015tka,Lozano:2016kum,Lozano:2016wrs}. This is the Abelian T-dual (ATD) of AdS$_3\times$S$^3\times$CY$_2$ along the Hopf-fibre of the S$^3$, and orbifolds thereof, that also preserve (0,4) of the supersymmetries of the original D1-D5 system. The orbifold solutions describe the D1-D5-KK system, and are dual to (0,4) CFTs that have been discussed in the literature  \cite{Kutasov:1998zh,Sugawara:1999qp,Larsen:1999dh,Okuyama:2005gq,Kim:2015gha,Hanany:2018hlz}.

\subsection{The NATD solution}

 The non-Abelian T-dual (NATD) of AdS$_3\times$S$^3\times$T$^4$ with respect to a freely acting  SU(2) subgroup of its SO(4) R-symmetry group was constructed in \cite{Sfetsos:2010uq}. As in other NATD examples, the space dual to S$^3$ becomes, locally, $\mathbb{R}\times \text{S}^2$. The SO(4) R-symmetry is reduced to an SU(2) R-symmetry, and the solution is rendered (0,4) supersymmetric \cite{Lozano:2015bra}. Due to our lack of knowledge of how non-Abelian T-duality extends beyond spherical worldsheets  \cite{Alvarez:1993qi}, the space is globally unknown. In this section we will resort to holography in order to construct a compact internal space for which a well-defined 2d dual CFT exists, following the strategy in \cite{Lozano:2016kum,Lozano:2016wrs,Lozano:2017ole,Itsios:2017cew,Lozano:2018pcp}.

We start generalising the solution constructed in \cite{Sfetsos:2010uq} to arbitrary D1 and D5 brane charges and a compact CY$_2$ four dimensional internal space. The most general solution reads 
\begin{eqnarray}
ds_{10}^2&=&4L^2 ds^2(\text{AdS}_3)+M^2 ds^2(\text{CY}_2)+4L^2 ds^2(\text{S}^3)\\
e^{2\Phi}&=&1\\
F_3&=&8L^2\text{vol}(\text{S}^3)\\
F_7&=&-8L^2M^4\text{vol}(\text{S}^3)\wedge\text{vol}(\text{CY}_2)\, .
\end{eqnarray}
The corresponding D1 and D5 brane charges are given by
\begin{eqnarray}
N_1&=&\frac{1}{(2\pi)^6}\int_{\text{S}^3\times \text{CY}_2}F_7=\frac{4L^2M^4}{(2\pi)^4}\text{Vol}(\text{CY}_2)\\
N_5&=&\frac{1}{(2\pi)^2}\int_{\text{S}^3} F_3=4L^2
\end{eqnarray}

The NATD with respect to a freely acting SU(2) group on the S$^3$ reads
\begin{eqnarray}
ds_{10}^2&=&4L^2 ds^2(\text{AdS}_3)+M^2 ds^2(\text{CY}_2)+\frac{d\rho^2}{4L^2}+\frac{L^2\rho^2}{4L^4+\rho^2}ds^2(\text{S}^2) \\
e^{2\Phi}&=&\frac{4}{4L^6+L^2\rho^2}\\
\label{B2NATD1}
B_2&=&-\frac{\rho^3}{2(4L^4+\rho^2)}\text{vol}(\text{S}^2)\\
F_0&=&L^2\\
F_2&=&-\frac{L^2\rho^3}{2(4L^4+\rho^2)}\text{vol}(\text{S}^2)\\
F_4&=&-L^2 M^4 \text{vol}(\text{CY}_2) \\
F_6&=& \frac{L^2 M^4 \rho^3}{2(4L^4+\rho^2)}\text{vol}(\text{CY}_2)\wedge \text{vol}(\text{S}^2)
\end{eqnarray}
It is easy to see that this solution fits locally in the class of AdS$_3\times$S$^2\times$CY$_2$ solutions constructed in \cite{LMNR1}, with the simple choices 
\begin{eqnarray}
u&=& 4L^4 M^2 \rho\label{uNATD}\\
h_4&=& L^2 M^4 \rho \label{h4NATD} \\
h_8&=&F_0\rho\, .
\end{eqnarray}
These functions define a regular, albeit non-compact, solution. We will shortly be discussing various possibilities that define it globally. For now let us analyse the associated quantised charges.
   
We start discussing the relevance of large gauge transformations. Close to $\rho=0$ the 3d transverse space is $\mathbb{R}^3$, while for large $\rho$ it is $\mathbb{R}\times \text{S}^2$.
This implies that for finite $\rho$ there is a non-trivial S$^2$ on which we can compute $\int_{\text{S}^2}B_2$, which needs to satisfy
\begin{equation}
\label{fundreg}
\frac{1}{4\pi^2}|\int_{\text{S}^2} B_2|\in[0,1).
\end{equation}
For $B_2$ as in  (\ref{B2NATD1}) this implies that  
a large gauge transformation needs to be performed as we move in $\rho$, such that $B_2\rightarrow B_2+n\pi \text{vol}_{S^2}$ for $\rho\in [\rho_n,\rho_{n+1}]$, with 
\begin{equation}
\label{rhon}
\frac{\rho_n^3}{4L^4+\rho_n^2}=2n\pi.
\end{equation} 
The non-compactness of $\rho$ is then reflected in the existence of large gauge transformations of infinite gauge parameter $n$. Moreover, taking into account large gauge transformations, we see that even if the 2-form and 6-form Page fluxes vanish identically,
\begin{equation}
{\hat f}_2=F_2-F_0\wedge B_2=0, \qquad {\hat f}_6=F_6-B_2\wedge F_4=0,
\end{equation}
implying the absence of D6 and D2 brane quantised charges, there is a non-zero contribution when $n\neq 0$, such that
\begin{eqnarray}
N_{8}&=&2\pi F_0=2\pi L^2\\
N_{6}&=&\frac{F_0}{2\pi}n\pi \text{Vol}(\text{S}^2)=n N_{8}\\
N_{4}&=&\frac{1}{(2\pi)^3}\int_{\text{CY}_2}F_4=\frac{L^2 M^4}{(2\pi)^3}\text{Vol}(\text{CY}_2)\\
N_{2}&=&\frac{1}{(2\pi)^5} \int_{\text{CY}_2}F_4 n\pi \text{Vol}(S^2)=nN_{4}\\
N_{5}&=&\frac{1}{(2\pi)^2}\int_{\rho_n}^{\rho_{n+1}}\int_{S^2}H_3=1.
\end{eqnarray}

These conserved charges suggest that the D1-D5 system that underlies the Type IIB AdS$_3\times$S$^3\times$CY$_2$ solution has been mapped under the NATD transformation onto a brane system consisting on $n$ D2-D6 branes at each $[\rho_n,\rho_{n+1})$ interval, dissolved in a D4-D8 bound state, due to the non-vanishing $B_2$-charge.  The corresponding brane distribution is depicted in table \ref{D2-NS5}. 
\begin{table}[ht]
	\begin{center}
		\begin{tabular}{| l | c | c | c | c| c | c| c | c| c | c |}
			\hline		    
			& 0 & 1 & 2 & 3 & 4 & 5 & 6 & 7 & 8 & 9 \\ \hline
			D2 & x & x & &  &  &  & x  &   &   &   \\ \hline
			D4 & x & x &  &  &  &   & x &  & x & x  \\ \hline
			D6 & x & x & x & x & x & x & x  &   &   &   \\ \hline
			D8 & x & x &x  & x & x &  x & x &  & x & x  \\ \hline
			NS5 & x & x &x  & x & x & x  &   &   &  &  \\ \hline
		\end{tabular} 
	\end{center}
	\caption{Distribution of branes compatible with the quantised charges of the NATD solution. $(y^0,y^1)$ are the directions where the 2d CFT lives, $(y^2,\dots,y^5)$ parameterise the CY$_2$, $y^6=\rho$, $y^7$ is the radius of AdS$_3$ and $(y^8,y^9)$ span the S$^2$.}   
	\label{D2-NS5}	
\end{table} 
This  configuration is the same as the one underlying the solutions constructed in \cite{Dibitetto:2017klx}, and, as in that case, it can be related to the $\frac18$-BPS brane set-up depicted in table \ref{D6-NS5-D8-D2-D4-first}, where the SU(2)$_R$ symmetry is manifest, through a rotation in the $(x^6,x^7)$ subspace. Due to the non-compactness of $\rho$ the brane system is however infinite. This suggests a relation with the linear quiver with infinite gauge groups discussed in section \ref{linear-quiver}, that we can now make more explicit.

Indeed, given that $h_4$ and $u$, as given by  (\ref{h4NATD}) and (\ref{uNATD}), satisfy the condition $h_4\sim u$, the NATD solution fits in the class of solutions that can be related to AdS$_7$ solutions, discussed in section 3. Both solutions are related explicitly through the mapping
\begin{equation}
u=162 F_0 L^4\rho\, , \qquad P=\frac{2\sqrt{3}}{\pi}L^2,
\end{equation}
with $P$ as introduced in  (\ref{bati}). This selects the NATD solution with $M^2=\frac{3^4}{2}L^2$ \footnote{This restriction is imposed because the AdS$_7$ solution depends on one single parameter, $P$, while a generic NATD solution depends on two parameters, $L$ and $M$.},
as the one related to the 6d (1,0) linear quiver discussed in section \ref{linear-quiver}.  These relations show that in the supergravity limit $L>>1$ the D6-branes are sent off to infinity. In this
 way we can think of the NATD solution as the {\it leading order} in an expansion in $P$, of the AdS$_7$ solution dual to the 6d linear quiver with gauge groups of increasing ranks, terminated with flavour D6-branes.
 
 In the next subsections we define other ways of completing the NATD solution with compact AdS$_3$ solutions. This will be valid for arbitrary values of the charges.

\subsection{2d (0,4) dual CFT}

As we have seen, the quantised charges of the NATD solution are compatible with an infinite brane system consisting on D2 and D6 branes stretched between NS5 branes. The D6 branes are wrapped on the CY$_2$, and thus share the same number of non-compact directions of the D2 branes.

General 2d (0,4) quiver theories associated to the 1/8-BPS D2-D4-D6-D8-NS5 brane configurations depicted in table \ref{D6-NS5-D8-D2-D4-first} were constructed in \cite{LMNR2}. For the particular configuration corresponding to the NATD solution the quiver contains two infinite families of nodes, associated to D2 and wrapped D6 branes, with gauge groups of increasing ranks, and no flavours. This quiver is depicted in figure  \ref{infinite-quiver}. We next summarise its main ingredients (the reader can find more details in reference \cite{LMNR2}):
\begin{figure}
\centering
\includegraphics[scale=0.9]{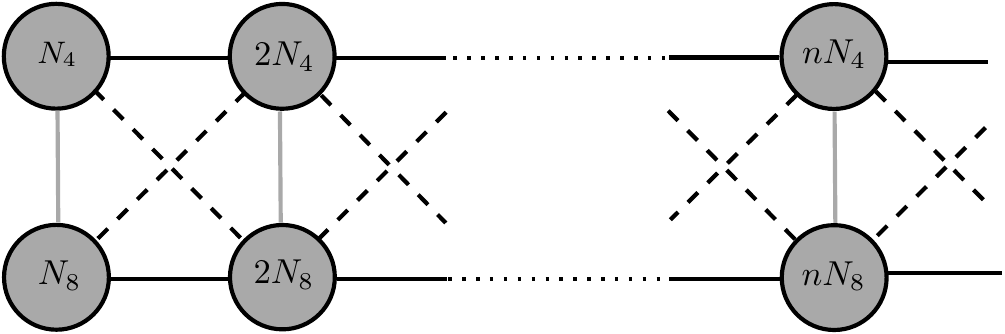}
\caption{Infinite quiver associated to the NATD solution.}
\label{infinite-quiver}
\end{figure} 

\begin{itemize}
\item To each gauge node corresponds a (0,4) vector multiplet plus a (0,4) twisted hypermultiplet in the adjoint representation of the gauge group. In terms of (0,2) multiplets, the first consists on a vector multiplet and a Fermi multiplet in the adjoint, and the second to two chiral multiplets forming a (0,4) twisted hypermultiplet, also in the adjoint. The (0,4) vector and the (0,4) twisted hypermultiplet combine to form a (4,4) vector multiplet. They are represented by circles.
\item Between each pair of horizontal nodes there are two (0,2) Fermi multiplets, forming a (0,4) Fermi multiplet, and two (0,2) chiral multiplets, forming a (0,4) hypermultiplet, each in the bifundamental representation of the gauge groups. The (0,4) Fermi multiplet and the (0,4) hypermultiplet combine into a (4,4) hypermultiplet. They are represented by black solid lines.
\item Between each pair of vertical nodes there are two (0,2) chiral multiplets forming a (0,4) hypermultiplet, in the bifundamental representation of the gauge groups. They are represented by grey solid lines.
\item Between each gauge node and any successive or preceding node there is one (0,2) Fermi multiplet in the bifundamental representation. They are represented by dashed lines.
\item Between each gauge node and a global symmetry node there is one (0,2) Fermi multiplet in the fundamental representation of the gauge group. They are again represented by dashed lines.
\end{itemize}

Note that the resulting quiver, depicted in figure \ref{infinite-quiver}, can be divided into two, horizontal, (4,4) linear quivers consisting on (4,4) gauge groups with increasing ranks connected by (4,4) bifundamental hypermultiplets. They correspond to the two (4,4) D6-NS5-D8 and D2-NS5-D4 {\it subsectors} of the brane configuration. The coupling between these two linear quivers through (0,4) hypermultiplets and (0,2) Fermi multiplets renders however the complete quiver (0,4) supersymmetric (see \cite{LMNR2} for more details).

The previous fields contribute to the gauge anomaly of a generic $\text{SU}(N_i)$ gauge group as:
\begin{itemize}
\item A (0,2) vector multiplet contributes with a factor of $-N_i$.
\item A (0,2) chiral multiplet in the adjoint representation contributes with a factor of $N_i$.
\item A (0,2) chiral multiplet in the bifundamental representation contributes with a factor of $\frac12$.
\item A (0,2) Fermi multiplet in the adjoint representation contributes with a factor of $-N_i$.
\item A (0,2) Fermi multiplet in the fundamental or bifundamental representation contributes with a factor of $-\frac12$.
\end{itemize}
Following these rules it is easy to see that the coefficient of the anomalous correlator of the symmetry currents $<J^A_{\mu}(x)J^B_{\nu}(x)>\sim k\delta_{\mu\nu}\delta^{AB}$ vanishes for each gauge group (see \cite{LMNR2} for more details) - hence the gauge anomalies vanish. By assigning R-charges to the different multiplets (see \cite{LMNR2} for the
precise assignation), we can calculate the U(1)$_R$ anomaly (for U(1)$_R$
inside SU(2)$_R$). The correlation function $<j_\mu(x)j_\nu(y)>$ for two
U(1)$_R$ currents is proportional to the number of ${\cal N}=(0,4)$
hypermultiplets minus the number of ${\cal N}=(0,4)$ vector multiplets.
This result is conserved when flowing to lower energies. In the far IR,
when the theory is proposed to become conformal the R-symmetry anomaly is
related to the central charge as indicated below.

\subsubsection{Central charge}

\begin{figure}
\centering
\includegraphics[scale=0.9]{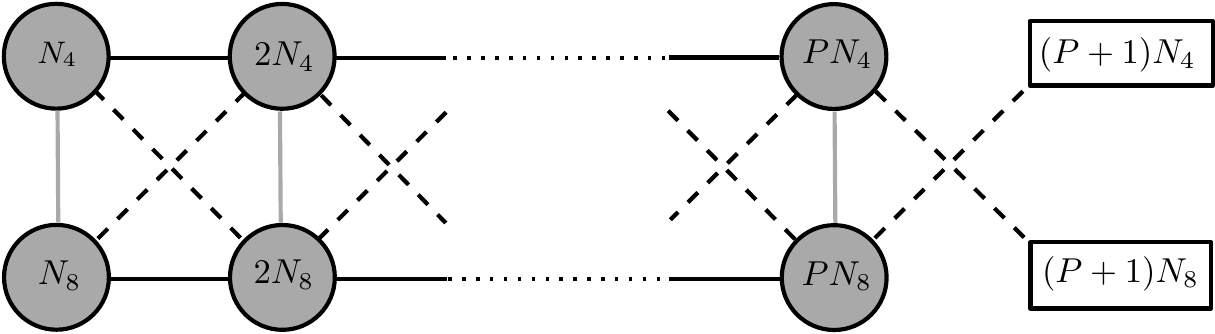}
\caption{Completed quiver with a finite number of gauge groups.}
\label{infinite-quiver-completed}
\end{figure}

Let us now discuss the central charge associated to this quiver. We compute it using the formula (see \cite{LMNR2,Putrov:2015jpa})
\begin{equation}
c=6(n_{hyp}-n_{vec}),
\end{equation}
where $n_{hyp}$ counts the number of fundamental and bifundamental hypermultiplets and $n_{vec}$ of vector multiplets. Clearly, these numbers are infinite for our quiver in figure \ref{infinite-quiver}. However, since they are subtracted in the computation of the central charge, they could still render a finite value. Terminating the space at a given $n=P$ and analysing the behaviour when $P$ goes to infinity we show however that this is not the case. Anomaly cancellation enforces that flavour groups must be added to both gauge groups at the end of the quiver. The resulting quiver is the one shown in figure
 \ref{infinite-quiver-completed}. This quiver was discussed in \cite{LMNR2}, as one of the anomaly free examples analysed therein.  For completeness we reproduce here the computation of its central charge.

The hypermultiplets that contribute to the counting of $n_{hyp}$ are the two chiral multiplets in each solid horizontal line, plus the two chiral multiplets in each vertical line. They give
\begin{equation}
n_{hyp}=\sum_{j=1}^{P-1} j(j+1) (N_4^2+N_8^2)+\sum_{j=1}^P j^2 N_4 N_8 =(N_4^2+N_8^2)(\frac{P^3}{3}-\frac{P}{3})+N_4N_8 (\frac{P^3}{3}+\frac{P^2}{2}+\frac{P}{6})
\end{equation}
Vector multiplets come from each node in the quiver, such that:
\begin{equation}
n_{vec}=\sum_{j=1}^{P} (j^2 N_4^2-1 +j^2 N_8^2 -1)=(N_4^2+N_8^2)(\frac{P^3}{3}+\frac{P^2}{2}+\frac{P}{6})-2P
\end{equation}
This gives for the central charge
\begin{equation}
c=6\Bigl[-(N_4^2+N_8^2)(\frac{P^2}{2}+\frac{P}{2})+N_4N_8(\frac{P^3}{3}+\frac{P^2}{2}+\frac{P}{6})+2P\Bigr].
\end{equation}
To leading order in $P$ we have,
\begin{equation}
\label{fieldtheoryc}
c\sim 2N_4N_8P^3.
\end{equation}
The central charge thus diverges with $P^3$ for the infinite quiver dual to the NATD solution. Still, it is useful to show that (\ref{fieldtheoryc}) coincides with the holographic central charge for $\rho\in [0,\rho_P]$, with $\rho_P$ satisfying (\ref{rhon}). Note that for large $P$ we can simply take $\rho_P=2\pi P$. Using (\ref{cads3}) we find 
 for $\rho\in [0,2\pi P]$,
\begin{equation}
c_{hol}=\frac{\pi}{2 G_N}(2\pi)^5 N_4 N_8 P^3=2N_4N_8 P^3,
\end{equation}
in agreement with the field theory result.

Our calculation shows the precise way in which the central charge diverges due to the non-compact field theory direction. It also gives us a possible way to regularise the infinite CFT dual to the NATD solution. Indeed, the quiver depicted 
 in figure \ref{infinite-quiver-completed} describes a well-defined 2d (0,4) CFT, that can be used to find a global completion of the non-Abelian T-dual solution. This completion is obtained glueing the non-Abelian T-dual solution at $\rho_P=2\pi P$ to another solution in \cite{LMNR1} that terminates the space at 
 $\rho=2\pi (P+1)$. We present the details of this completion in the next subsection. In section \ref{glueing2} we present a different completion, which makes manifest that this procedure is not unique and that one can device different global completions of the NATD solution, as stressed in 
\cite{Lozano:2016kum}.



\subsection{Completions} \label{completions}

In this section we present two possible completions of the NATD solution. The AdS$_3$ example is particularly useful in this respect, because the completed solution is not only explicit but also extremely simple, as opposed to other examples in higher dimensions  \cite{Lozano:2016kum,Lozano:2016wrs,Itsios:2017cew}.

\subsubsection{Completion with O-planes} \label{Oplanes}

The simplest way to complete the NATD solution is by terminating the infinite linear quiver at a certain value of $\rho$, as we have done in the previous subsection. 
We take this to be $\rho=2\pi (P+1)$, with $P\in \mathbb{Z}$, and choose the $u$, $h_8$ and $h_4$ functions such that:
\begin{equation}
u= 4L^4 M^2 \rho, \qquad 0 \leq \rho \leq 2\pi (P+1)
\end{equation}
 \begin{equation} \label{h8symmetric}
h_8(\rho)
                    =F_0 . \left\{ \begin{array}{ccrcl}
                        \rho & \quad 0\leq \rho\leq 2\pi P \\
                       P(2\pi (P+1)-\rho) & \quad 2\pi P \leq \rho \leq 2\pi (P+1). 
                                             \end{array}
\right.
\end{equation}

 \begin{equation} \label{h4symmetric}
h_4(\rho)
                    =L^2 M^4 . \left\{ \begin{array}{ccrcl}
                        \rho & \quad 0\leq \rho\leq 2\pi P \\
                       P(2\pi (P+1)-\rho) & \quad 2\pi P \leq \rho \leq 2\pi (P+1). 
                                             \end{array}
\right.
\end{equation}

The explicit form of the metric, dilaton and fluxes in the $2\pi P \leq \rho \leq 2\pi (P+1)$ region can be found in Appendix \ref{completionOplane}. One can check that the NS sector is continuous at $\rho=2\pi P$. The 2-form and 6-form Page fluxes are also continuous once large gauge transformations are taken into account. They are given by
\begin{eqnarray} 
&&{\hat f}_2
                    =-F_0\, {\rm vol}(\text{S}^2) . \left\{ \begin{array}{ccrcl}
                         n\pi & \quad 0\leq n\leq P \\
                        P \pi (P+1-n)& \quad P \leq n \leq P+1
                                             \end{array}
\right.\\
&&{\hat f}_6
                    =L^2 M^4\, {\rm vol}(\text{S}^2)\wedge {\rm vol}(\text{CY}_2) . \left\{ \begin{array}{ccrcl}
                        n\pi & \quad 0\leq n\leq P \\
                        P \pi (P+1-n)& \quad P \leq n \leq P+1,
                                                                    \end{array}
\right.
\end{eqnarray}
so they vanish at $n=P+1$, where the geometry terminates. We show below that at this point the background has a singularity associated to O6-O2 planes. In turn there is a discontinuity in $F_0$ and $F_4$ at $n=P$ that is translated into $(P+1) N_8$ and $(P+1)N_4$ additional flavours connected to the nodes corresponding to $PN_4$ D2 and $PN_8$ D6 branes, respectively. This is exactly as in the quiver depicted in figure \ref{infinite-quiver-completed}.

The expressions of the metric and dilaton in the $2\pi P\leq \rho \leq 2\pi (P+1)$ region, given by equations (\ref{O1}), (\ref{O2}) in Appendix \ref{completionOplane}, show that close to $\rho=2\pi (P+1)$ they  behave as 
\begin{equation}
ds^2\sim x^{-1}ds^2(\text{AdS}_3)+M^2 ds^2(\text{CY}_2)+x \big(dx^2+ds^2(\text{S}^2)\big), \qquad e^{2\phi}\sim x^{-1}
\end{equation}
where $x=\rho-2(P+1)$. This singular behaviour corresponds to the intersection of an O6 fixed plane lying on AdS$_3\times$CY$_2$ with O2-planes lying on AdS$_3$ and smeared on CY$_2\times$S$^2$. Even if it is not clear what this object is in string theory, the fact that the solution has a well-defined dual CFT suggests that it should be possible to give it a meaning. 

\subsubsection{Glueing the NATD to itself}\label{glueing2}

Another interesting way of defining globally the NATD solution is by glueing it to itself. In this case we take:
 \begin{equation} \label{u2symmetric}
u(\rho)
                    =4L^4 M^2 , \quad 0\leq \rho\leq 4\pi P. 
                    \end{equation}

 \begin{equation} \label{h8symmetric}
h_8(\rho)
                    =F_0 . \left\{ \begin{array}{ccrcl}
                        \rho & \quad 0\leq \rho\leq 2\pi P \\
                       4\pi P-\rho & \quad 2\pi P \leq \rho \leq 4\pi P. 
                                             \end{array}
\right.
\end{equation}

 \begin{equation} \label{h4symmetric}
h_4(\rho)
                    =L^2 M^4 . \left\{ \begin{array}{ccrcl}
                        \rho & \quad 0\leq \rho\leq 2\pi P \\
                       4\pi P-\rho & \quad 2\pi P \leq \rho \leq 4\pi P. 
                                             \end{array}
\right.
\end{equation}

\begin{figure}
\centering
\includegraphics[scale=0.8]{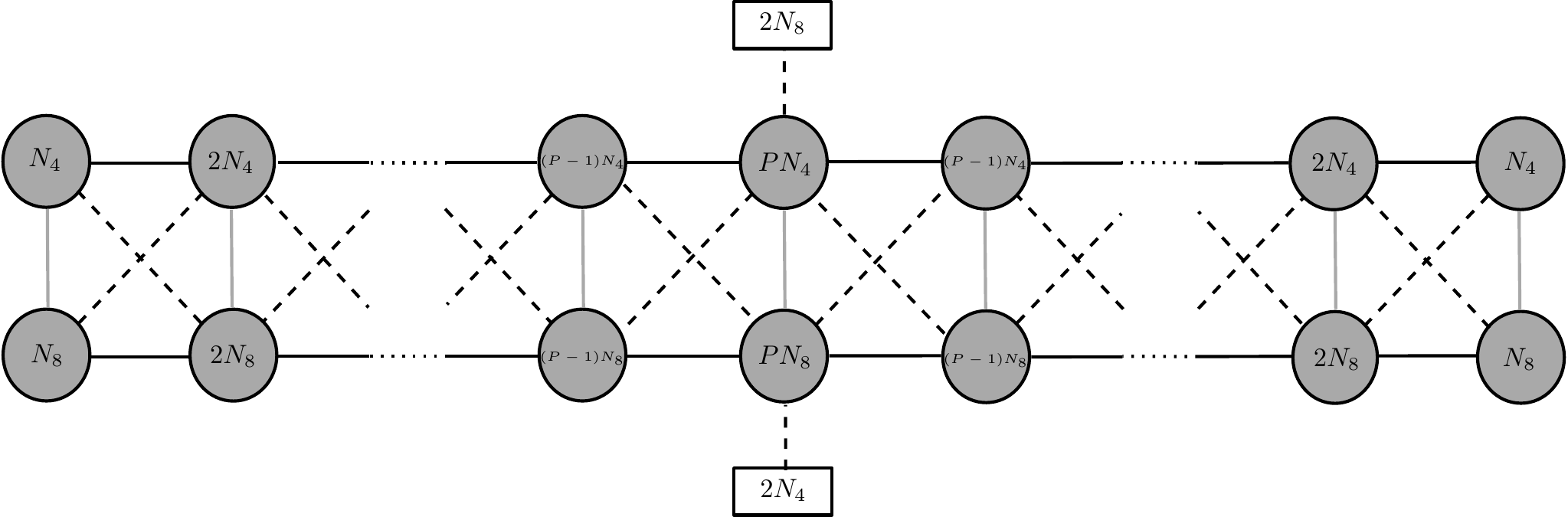}
\caption{Symmetric quiver associated to the NATD solution glued to itself.}
\label{symmetric-quiver}
\end{figure}

The explicit form of the metric, dilaton and fluxes in the $2\pi P \leq \rho \leq 4\pi P$ region can be found in Appendix \ref{completionOplane}. One can check that the NS sector is continuous at $\rho=2\pi P$. The 2-form and 6-form Page fluxes are also continuous once large gauge transformations are taken into account. They are given by
\begin{equation} 
{\hat f}_2
                    =-F_0\, {\rm vol}(\text{S}^2). \left\{ \begin{array}{ccrcl}
                         n\pi & \quad 0\leq n\leq P \\
                        (2P-n) \pi & \quad P \leq n \leq 2P
                                             \end{array}
\right.
\end{equation}
and
\begin{equation} 
{\hat f}_6
                    =L^2 M^4\, {\rm vol}(\text{S}^2)\wedge {\rm vol}(\text{CY}_2). \left\{ \begin{array}{ccrcl}
                        n\pi & \quad 0\leq n\leq P \\
                       (2P-n) \pi & \quad P \leq n \leq 2P
                                             \end{array}
\right.
\end{equation}
Therefore, they are both continuous at $n=P$ and vanish at $n=2P$. The corresponding quantised charges are:
\begin{equation} 
N_6
                    =\left\{ \begin{array}{ccrcl}
                        n N_8& \quad 0\leq n\leq P \\
                        (2P-n) N_8 & \quad P \leq n \leq 2P
                                             \end{array}
\right.
\end{equation}
and
\begin{equation} 
N_2
                    =\left\{ \begin{array}{ccrcl}
                        n N_4 & \quad 0\leq n\leq P \\
                       (2P-n) N_4 & \quad P \leq n \leq 2P
                                             \end{array}
\right.
\end{equation}
where $N_6$ denotes anti-D6 brane charge, $N_2$ D2-brane charge and $N_8=\pm 2\pi F_0$ in the two regions. For $N_4$ we have 
\begin{equation}
N_4=\frac{1}{(2\pi)^3}\int {\hat f}_4=\frac{1}{(2\pi)^3}\int F_4=\mp  \frac{L^2 M^4}{(2\pi)^3}\text{Vol}(\text{CY}_2)
\end{equation} 
in the two regions. Thus, the D2 and D6 brane charges increase linearly in the $0\leq n\leq P$ region, corresponding to the NATD solution, and decrease linearly in the $P\leq n\leq 2P$ region, till they vanish at $n=2P$, where the geometry terminates. At this point the S$^2$ shrinks smoothly. The discontinuity of $N_8$ and $N_4$ at $n= P$ is translated into $2 N_8$ and $2 N_4$ additional flavours at the nodes with flavour groups $PN_4$ and $PN_8$, respectively. 
The associated quiver is the one depicted in figure \ref{symmetric-quiver}. The $2N_8$ and $2N_4$ flavour groups contribute each with one (0,2) Fermi multiplet in the fundamental representation of the corresponding gauge group. As for the quivers constructed in \cite{LMNR2}, the flavour group introduced at the node associated to D2-branes arises from D8-branes while that introduced at the node associated to D6-branes arises from D4-branes.

The central charge of this quiver is given by
\begin{equation}
c=6\Bigl[(N_4^2+N_8^2)(-P)+N_4N_8(\frac23 P^3+\frac{P}{3})+4P-2\Bigr].
\end{equation}
To leading order in $P$ this gives
\begin{equation}
c=4N_4N_8P^3,
\end{equation}
which one can check is in agreement with the holographic central charge.

\subsection{The Abelian T-dual limit}

The non-Abelian T-dual solution defined in $\rho\in [\rho_n,\rho_{n+1}]$ gives rise to the Abelian T-dual, along the Hopf-fibre of the S$^3$, of the original AdS$_3\times$S$^3\times$CY$_2$ background, in the limit in which $n$ goes to infinity \cite{Macpherson:2015tka,Lozano:2016kum,Lozano:2016wrs}. In this subsection we will be interested in the ATD solution, and orbifolds thereof, in its own right, as another explicit example in the class in \cite{LMNR1}. 

The ATD solution is given by
\begin{eqnarray}
ds_{10}^2&=&4L^2 ds^2(\text{AdS}_3)+M^2 ds^2(\text{CY}_2)+\frac{d\psi^2}{4L^2}+L^2 ds^2(\text{S}^2) \\
e^{2\Phi}&=&\frac{4}{L^2}\\
\label{B2NATD}
B_2&=&-\frac{\psi}{2}\text{vol}(\text{S}^2)\\
F_2&=&-\frac{L^2}{2}\text{vol}(\text{S}^2)\\
F_6&=& \frac12 M^4 L^2 \text{vol}(\text{CY}_2)\wedge \text{vol}(\text{S}^2),
\end{eqnarray}
where $\psi$ is the ATD of the Hopf-fibre direction, normalised such that $\psi\in [0,2\pi]$. Upon dualisation, the (4,4) supersymmetries of the original solution are reduced to (0,4) \cite{Lozano:2015bra}, and the solution fits in the classification in \cite{LMNR1}. The corresponding $u$, $h_4$ and $h_8$ functions are given by
\begin{eqnarray}
u&=&4L^4 M^2\\
h_4&=&L^2 M^4\\
h_8&=&L^2.
\end{eqnarray}
The quantised charges are,
\begin{equation}
N_2=\frac{L^2 M^4}{(2\pi)^4}\text{Vol}(\text{CY}_2)\, , \qquad N_6=L^2\, , \qquad N_5=1
\end{equation}
so using (\ref{cads3}) the holographic central charge gives
\begin{equation}
c_{hol}=6 N_2 N_6.
\end{equation}
One can check that this is reproduced from the NATD solution for $\rho\in [\rho_n,\rho_{n+1}]$ and $n$ large, using that $N_2=nN_4$ and $N_6=nN_8$ in this interval. 
The brane set-up describing the ATD solution consists on $N_2$ D2-branes and $N_6$ D6-branes, wrapped on the CY$_2$, stretched along the $\psi$ circular direction between two NS5-branes that are identified. 

\begin{figure}
\centering
\includegraphics[scale=0.8]{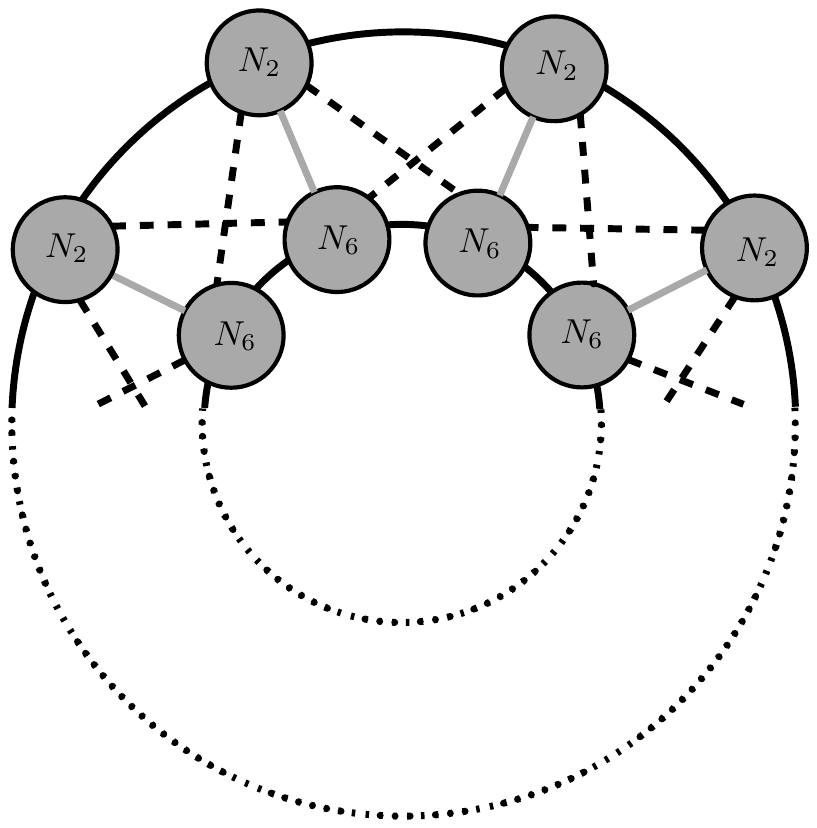}
\caption{Circular quiver associated to the (orbifolded) ATD solution.}
\label{ATD}
\end{figure}

Orbifolds of this solution can be constructed taking $\psi\in [0,2\pi N]$. They are T-dual to the AdS$_3\times$S$^3/\mathbb{Z}_N\times$CY$_2$ solution in Type IIB that describes the D1-D5-KK system \cite{Kutasov:1998zh,Sugawara:1999qp,Larsen:1999dh,Okuyama:2005gq,Kim:2015gha}. The Type IIA brane realisation of this system is depicted in figure \ref{ATD}. From this quiver we have that

\begin{equation}
n_{vec}=(N_2^2-1+N_6^2-1)N, \qquad n_{hyp}=(N_2^2+N_6^2+N_2N_6)N.
\end{equation}
One then obtains a central charge
\begin{equation}
c=6(n_{hyp}-n_{vec})=6N_2N_6 N+12 N.
\end{equation}
This gives in the large $N_2, N_6$ limit,
\begin{equation}
c\sim 6N_2N_6 N,
\end{equation}
in agreement with the central charge of the D1-D5-KK system \cite{Kutasov:1998zh}\footnote{This central charge was computed using the Brown-Henneaux formula \cite{Brown:1986nw}. One can also use  (\ref{cads3}), which generalises the central charge therein to non-trivial warping and dilaton.}.

\begin{figure}
\centering
\includegraphics[scale=0.8]{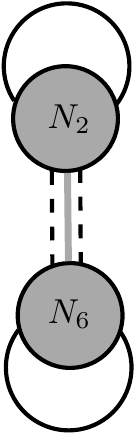}
\caption{Quiver associated to the ATD solution.}
\label{ATDII}
\end{figure}

For $N=1$ the quiver in figure \ref{ATD} reduces to the quiver depicted in figure \ref{ATDII}. 
The (4,4) hypermultiplets connecting $N_2$ nodes and $N_6$ nodes among themselves become (4,4) hypermultiplets in the adjoint representation.  In turn, the (0,2) Fermi multiplets connecting each $N_2$ ($N_6$) node with adjacent $N_6$ ($N_2$) nodes combine into (0,4) Fermi multiplets connecting each $N_2$ node with its respective $N_6$ node, which together with the (0,4) hypermultiplets between them give (4,4) hypermultiplets in the bifundamental.
In this way supersymmetry is enhanced to (4,4), and the quiver describes the D1-D5 system in terms of the D2 and D6-brane charges of the Abelian T-dual solution\footnote{See \cite{David:2002wn}, section 4, for this analysis in Type IIB.}.

 \section{Relation with the AdS$_3\times$S$^2$ flows of Dibitetto-Petri}
\label{AdS3flow}

 In \cite{Dibitetto:2017tve,Dibitetto:2017klx} Dibitetto and Petri (DP) constructed various BPS flows within minimal $\mathcal{N}=1$ 7d supergravity that are asymptotically locally AdS$_7$. These flows are described by warped AdS$_3$ solutions  triggered by a non-trivial dyonic 3-form potential. A particularly interesting solution was constructed in \cite{Dibitetto:2017tve}, which was shown to interpolate between asymptotically locally AdS$_7$ and AdS$_3 \times$T$^4$ geometries. The UV AdS$_7$ limit is (asymptotically locally) the reduction to 7d of the AdS$_7$ solutions of massive IIA constructed in \cite{Apruzzi:2013yva}. In this subsection we would like to explore the 10d uplift of the IR AdS$_3\times$T$^4$ limit, in connection with the subclass of solutions discussed in section 2, in the case in which CY$_2=$T$^4$.
 
The AdS$_3$ solution constructed in \cite{Dibitetto:2017tve} reads (see Appendix \ref{DPdetails} for the details),
\begin{eqnarray}
\label{metricAdS7DP}
ds_7^2&=&e^{2U(r)}ds(\text{AdS}_3)^2+e^{2V(r)}dr^2+e^{2W(r)}ds(\text{S}^3)^2,\nonumber\\
X&=&X(r),\nonumber\\
B_{(3)}&=&k(r) \text{vol}(\text{AdS}_3)+l(r) \text{vol}(\text{S}^3),
\end{eqnarray}
where $X$, $U$, $V$, $W$, $k$ and $l$ are functions of $r$ discussed in the Appendix \ref{DPdetails}. This solution is 
asymptotically locally AdS$_7$ when $r\rightarrow\infty$ , while when $r\to 0$ it flows to an AdS$_3\times$T$^4$ non-singular limit, given by\footnote{As compared to  \cite{Dibitetto:2017tve}, we write the 7d metric in terms of an AdS$_3$ space of radius one.},
\begin{equation}
\label{metricr0}
ds_7^2=\frac{2^{31/5}}{g^2}\Bigl(\frac{3^{2/5}}{5^2}ds^2(\text{AdS}_3)+\frac{4}{3^{8/5}}ds^2(\text{T}^4)\Bigr),
\end{equation}
and 
\begin{equation}
B_3=-\frac12 \text{vol}(\text{AdS}_3)-4r^4 \text{vol}(\text{S}^3).
\end{equation}
As the AdS$_7$ asymptotic limit, this geometry is not a solution of 7d $\mathcal{N}=1$ minimal supergravity by itself, but rather the IR leading asymptotics of the flow. In the discussion that follows it will be useful to recall from Appendix \ref{DPdetails} that the values of the 7d scalar $X$ in the $r\to \infty$ and $r\to 0$ limits are $X=1$ and $X^5=2^{2}/3$, respectively.

7d ${\mathcal N}=1$ minimal supergravity can be consistently uplifted to massive IIA on a squashed S$^3$  \cite{Passias:2015gya}. Using the uplift formulae provided in Appendix \ref{DPdetails}, 
a family of AdS$_3$ solutions to massive IIA can thus be constructed from the DP flow.  This gives rise in the $r\to \infty$ limit to 10d geometries that asymptote to the AdS$_7 \times$S$^2 \times$I family of solutions in  \cite{Apruzzi:2013yva}. In turn, the geometry that is obtained in the 
AdS$_3\times$T$^4$ limit reads\footnote{Here we have taken $g^3=8\sqrt{2}$, which is the value for which the internal space and fluxes of the AdS$_7$ solutions in \cite{Apruzzi:2013yva} are recovered.} 
 \begin{eqnarray}
\label{ds10rto0-alpha}
ds_{10}^2&=&8\sqrt{2}\pi\sqrt{-\frac{\alpha}{\ddot{\alpha}}}\,\bigg(\frac{2^3 \sqrt{3}}{5^2}ds^2(\text{AdS}_{3})+\frac{2^5}{3\sqrt{3}}ds^2(\text{T}^4)\bigg)\nonumber\\
&&+\frac{2\sqrt{2}}{\sqrt{3}}\pi\sqrt{-\frac{\ddot{\alpha}}{\alpha}}dz^2+2\sqrt{6}\pi \frac{\alpha^{3/2}(-\ddot{\alpha})^{1/2}}{3\dot{\alpha}^2-8\alpha\ddot{\alpha}} ds^2(\text{S}^2)\\
e^{2\Phi}&=&2^33^8\sqrt{6}\pi^5\bigg(-\frac{\alpha}{\ddot{\alpha}}\bigg)^{3/2}\frac{1}{3\dot{\alpha}^2-8\alpha \ddot{\alpha}}\\
B_2&=&\pi \Bigl(-z+\frac{3\alpha\dot{\alpha}}{3\dot{\alpha}^2-8\alpha\ddot{\alpha}}\Bigr)\text{vol}(\text{S}^2) \\
F_2&=&\Bigl(\frac{\ddot{\alpha}}{162\pi^2}+\frac{3\pi F_0\alpha\dot{\alpha}}{3\dot{\alpha}^2-8\alpha\ddot{\alpha}}\Bigr)\text{vol}(\text{S}^2)\\
F_4&=&\frac{2^9}{3^4\pi}\bigg(\frac{\ddot{\alpha}}{5^3}\text{d}z\wedge\text{vol}(\text{AdS}_3)-\frac{2^5}{3^3}\dot{\alpha}\text{vol}(\text{T}^4)\bigg)\\
F_6&=&-\frac{2^{9}}{5^33^7}\frac{\alpha\ddot{\alpha}}{3\dot{\alpha}^2-8\alpha\ddot{\alpha}}\bigg(2^85^3\alpha\text{vol}(\text{T}^4)+3^4\dot{\alpha}\text{vol}(\text{AdS}_3)\wedge\text{d}z\bigg)\wedge\text{vol}(\text{S}^2) \label{ds10rto0-last}.
\end{eqnarray}
As in 7d, the uplift of the $r\to 0$ limit of the DP flow is not a solution to massive IIA by itself, but rather its IR leading asymptotics. We would like to see whether it can be completed by an AdS$_3\times$T$^4$ solution in the class of \cite{LMNR1}, with the same asymptotics. For that it is easy to realise that one can absorb the constant $X$ that causes the distortion of the internal space (we are referring to  (\ref{ds10a})-(\ref{F410a}) in Appendix \ref{DPdetails}) by simply modifying the mapping for the $h_4$ function in (\ref{ultima}) as $h_4= \frac{81}{8}X^5 u\leftrightarrow\frac{81}{8}X^5\alpha$. We then have for the IR geometry given by (\ref{ds10rto0-alpha})-(\ref{ds10rto0-last}),   
 \begin{eqnarray}
  \label{fieldthdir2}
  \rho&\leftrightarrow&2\pi z\\
  \label{ucuadrado2}
    u&\leftrightarrow&\alpha\\
    \label{h8alphaddot2}
    h_8&\leftrightarrow&-\frac{\ddot{\alpha}}{81 \pi^2} \\
    \label{ultima2}
   h_4&=& \frac{27}{2}u\leftrightarrow \frac{27}{2} \alpha\, .
    \end{eqnarray}
This gives for the AdS$_3\times$T$^4$ subspace
\begin{equation}
\frac{u}{\sqrt{h_4h_8}}ds^2(\text{AdS}_3)+\sqrt{\frac{h_4}{h_8}}ds^2(\text{T}^4) \leftrightarrow
\sqrt{6}\pi \sqrt{-\frac{\alpha}{\ddot{\alpha}}}\,\bigg(ds^2(\text{AdS}_{3})+\frac{3^3}{2}ds^2(\text{T}^4)\bigg).
\end{equation}
The result is a bonna fide AdS$_3\times$T$^4$ solution to massive IIA, supplemented with  $F_4$ and $F_6$ fluxes satisfying (\ref{fhat4}) and (\ref{fhat6}). The resulting 7d metric does not share however the asymptotics of the 7d metric arising from (\ref{ds10rto0-alpha}). Thus, the IR limit of the DP flow cannot be completed into an AdS$_3\times$T$^4$ solution in the class of \cite{LMNR1}, that shares its same asymptotics. This result excludes the RG flows constructed in \cite{Dibitetto:2017tve} as solutions interpolating between AdS$_3\times$T$^4$ geometries (in the subclass  defined in section 3) and the AdS$_7$ solutions constructed in  \cite{Apruzzi:2013yva}. Still, it should be possible to construct these flows, perhaps as $\mathbb{R}_{1,1}\times$CY$_2$ warped product geometries, as the ones discussed in \cite{Passias:2015gya}.


\section{Conclusions}

In this paper we have discussed some aspects of the class of AdS$_3\times$S$^2$ solutions with small ${\mathcal N}=(0,4)$ supersymmetry and SU(2)-structure constructed in \cite{LMNR1}. We have focused our analysis on a sub-set of solutions contained in  ``class I'' of \cite{LMNR1}, which are warped products of AdS$_3\times$S$^2\times$CY$_2$ over an interval with warpings respecting the symmetries of CY$_2$. 2d (0,4) CFTs dual to these solutions have been proposed recently in \cite{LMNRPRL,LMNR2}. 

We have established a map between the previous solutions and the AdS$_7$ solutions in \cite{Apruzzi:2013yva}, that allows one to interpret the former as duals to defects in 6d (1,0) CFTs. More precisely, the 2d dual CFT arises from wrapping on the CY$_2$ the  D6-NS5-D8 branes that underlie the AdS$_7$ solutions, and intersecting them with D2 and D4 branes. In this sense it combines {\it wrapped branes} and {\it defect branes}. The D2-branes are stretched between the NS5-branes, as the D6-branes, and the D4-branes are perpendicular, as the D8-branes. They give rise to (0,4) quivers with two families of gauge groups connected by matter fields \cite{LMNR2}. Each family is described by a (4,4) linear quiver and is connected with the other family by (0,4) and (0,2) multiplets, rendering the final quiver (0,4) supersymmetric. 

The previous mapping suggests that it should be possible to construct flows connecting the  AdS$_3\times$CY$_2$ solutions in the IR with the AdS$_7$ solutions in the UV. The presence of D2-D4 defects suggests that one should look at warped AdS$_3$ flows, as the ones discussed in \cite{Dibitetto:2017tve}, which interpolate between asymptotically locally AdS$_3\times$T$^4$ geometries, with an interpretation as 2d defect CFTs, and AdS$_7$ solutions. We have found however that our solutions have different asymptotics than the IR AdS$_3$ geometries considered in \cite{Dibitetto:2017tve}. This discrepancy could originate on the wrapped branes present in our solutions, more suggestive of an 
 $\mathbb{R}^{1,1}\times$CY$_2$ flow \cite{Bobev:2017uzs}, as the one constructed in \cite{Passias:2015gya}. It would be very interesting to find the explicit flow that interpolates between these two classes of solutions.

We have provided a thorough analysis of the AdS$_3\times$S$^2\times$CY$_2$ solution that arises from the Type IIB solution dual to the D1-D5 system through non-Abelian T-duality. Using the map between AdS$_3$ and AdS$_7$ solutions derived in the first part of the paper, we have {\it rediscovered} this solution as the leading order of the AdS$_7$ solution in the class in \cite{Apruzzi:2013yva} dual to a 6d linear quiver with gauge groups of increasing ranks, terminated by D6-branes. Secondly, we have provided two explicit global completions with AdS$_3$ solutions in the class in \cite{LMNR1}. One of these completions is obtained glueing the non-Abelian T-dual solution to itself, in a sort of orbifold projection around the point where the space terminates. This solution has a well-defined 2d dual CFT that we have studied. Orbifolds have previously played a role in the completion of NATD solutions, remarkably in the example discussed in \cite{Itsios:2017cew}, but this is the first time the explicit completed geometry has been constructed. 
The AdS$_3$ example provides indeed a very useful set-up where to test the role played by holography in extracting global information of NATD in string theory, following the ideas in \cite{Lozano:2016kum,Lozano:2016wrs,Lozano:2017ole,Itsios:2017cew,Lozano:2018pcp}.

\subsection*{Acknowledgements}
We would like to thank Giuseppe Dibitetto, Nicolo Petri, Alessandro Tomasiello and Stefan Vandoren for fruitful discussions. YL and AR are partially supported by the Spanish government grant PGC2018-096894-B-100 and by the Principado de Asturias through the grant FC-GRUPIN-IDI/2018/000174. NTM is funded by the Italian Ministry of Education, Universities and Research under the Prin project ``Non Perturbative Aspects of Gauge Theories and Strings'' (2015MP2CX4) and INFN. AR is supported by CONACyT-Mexico. We would like to acknowledge the Mainz Institute for Theoretical Physics (MITP) of the DFG Cluster of Excellence PRISMA$^{+}$ (Project ID 39083149) for its hospitality and partial support during the development of this work. YL and AR would also like to thank the Theory Unit at CERN for its hospitality and partial support during the completion of this work.

\appendix

\section{Completions of the NATD solution}
\label{completionOplane}

{\bf Completion with O-planes}

\noindent The metric, dilaton and fluxes of the NATD solution completed as indicated in section \ref{Oplanes} read, in the $2\pi P\leq \rho \leq 2\pi (P+1)$ region,
\begin{eqnarray}
ds^2&=&\frac{4L^2\rho}{P(2\pi(P+1)-\rho)}ds^2(\text{AdS}_3)+M^2 ds^2(\text{CY}_2)+\frac{P(2\pi(P+1)-\rho)}{4L^2\rho}d\rho^2\label{O1}\nonumber\\
&&+\frac{L^2 P\rho \, (2\pi(P+1)-\rho)}{4L^4+P^2(2\pi(P+1)-\rho)^2} ds^2(\text{S}^2) \\
e^{2\Phi}&=&\frac{4\rho}{L^2 P(2\pi(P+1)-\rho)\Bigl(4L^4+P^2(2\pi(P+1)-\rho)^2\Bigr)}\label{O2}\\
B_2&=&-\frac{\rho P^2(2\pi(P+1)-\rho)^2}{2\Bigl(4L^4+P^2(2\pi(P+1)-\rho)^2\Bigr)}{\rm vol}(\text{S}^2)\\
F_0&=&-PL^2\\
F_2&=&-\frac{L^2\left(P^3\left(2\pi\left(P+1\right)-\rho\right)^3+8L^4\pi P\left(P+1\right)\right)}{2\left(4L^4+P^2\left(2\pi\left(P+1\right)-\rho\right)^2\right)}\text{vol}(\text{S}^2)\\
F_4&=&L^2 M^4 P\, {\rm vol}(\text{CY}_2)
 \label{O8}
\end{eqnarray}

\noindent {\bf NATD solution glued to itself}

\noindent The metric, dilaton and fluxes of the NATD solution glued to itself read, in the $2\pi P\leq \rho \leq 4\pi P$ region,
\begin{eqnarray}
ds^2&=&\frac{4L^2\rho}{4\pi P-\rho}ds^2(\text{AdS}_3)+M^2 ds^2(\text{CY}_2)+\frac{4\pi P-\rho}{4L^2\rho}d\rho^2+\frac{L^2 \rho (4\pi P-\rho)}{4L^4+(4\pi P-\rho)^2} ds^2(\text{S}^2)\nonumber\\ \label{O12}\\
e^{2\Phi}&=&\frac{4\rho}{L^2 (4\pi P-\rho)\Bigl(4L^4+(4\pi P-\rho)^2\Bigr)}\label{O22}\\
B_2&=&-\frac{\rho (4\pi P-\rho)^2}{2\Bigl(4L^4+(4\pi P-\rho)^2\Bigr)}{\rm vol}(\text{S}^2)\\
F_0&=&-L^2\\
F_2&=&-\frac{L^2\Bigl((4\pi P-\rho)^3+16\pi PL^4\Bigr)}{2\Bigl(4L^4+(4\pi P-\rho)^2\Bigr)}
\text{vol}(\text{S}^2)\\
F_4&=&L^2 M^4 \, {\rm vol}(\text{CY}_2)
 \label{glueing}
\end{eqnarray}

\section{The Dibitetto-Petri flow in minimal ${\cal N}=1$ 7d supergravity} \label{DPdetails}

The solution discussed in section \ref{AdS3flow} was obtained in \cite{Dibitetto:2017tve} taking the following ansatz:
\begin{eqnarray}
\label{metricAdS7DP}
ds_7^2&=&e^{2U(r)}ds^2(\text{AdS}_3)+e^{2V(r)}dr^2+e^{2W(r)}ds^2(\text{S}^3),\nonumber\\
X&=&X(r),\nonumber\\
B_{(3)}&=&k(r) \text{vol}(\text{AdS}_3)+l(r) \text{vol}(\text{S}^3),
\end{eqnarray}
and vanishing vector fields.
Here $ds^2(\text{S}^3)$ is the metric of an S$^3$ with radius $\frac{2}{\kappa}$, parameterised as:
\begin{eqnarray}
\label{parametrizationS3}
e^1&=&\frac{1}{\kappa}d\theta_2,\nonumber\\
e^2&=&\frac{1}{\kappa}\cos{\theta_2} d\theta_3,\nonumber\\
e^3&=&\frac{1}{\kappa}(d\theta_1+\sin{\theta_2} d\theta_3),
\end{eqnarray}
and $ds^2(\text{AdS}_3)$ is the metric of an AdS$_3$ with radius $\frac{2}{L}$, parameterised as:
\begin{eqnarray}
\label{parametrizationAdS3}
e^1&=&\frac{1}{L}(dt-\sinh{x_1} dx^2),\nonumber\\
e^2&=&\frac{1}{L}dx^1,\nonumber\\
e^3&=&\frac{1}{L}\cosh{x_1}dx^2.
\end{eqnarray} 
$\text{vol}(\text{S}^3)$ and $\text{vol}(\text{AdS}_3)$ represent their corresponding volume forms.
DP showed that  (\ref{metricAdS7DP}) is a solution to minimal 7d sugra with $X$, $U$, $V$, $W$, $k$ and $l$ given by,
\begin{eqnarray}
\label{X}
X(r)&=&\frac{2^{2/5} h^{1/5}\left(-1+\rho ^8\right)^{2/5}}{\left(-8 L \rho ^4 \left(1+\rho ^8\right)+\sqrt{2} g \left(1+4 \rho ^4+4
\rho ^{12}+\rho ^{16}\right)\right)^{1/5}},\\
\label{U}
e^{2U(r)}&=&\frac{(\rho^4+1)^2}{4\rho^4 X^2},\\
\label{V}
e^{2V(r)}&=&\frac{4 X^8}{h^{2}},\\
\label{W}
e^{2W(r)}&=&\frac{(\rho^4-1)^2}{4\rho^4 X^2},\\
\label{l}
l(r)&=&\frac{1}{16 h \rho ^4 \left(\rho ^4+1\right)^2}[\sqrt{2} g \left(-1+4 \rho ^4+4 \rho ^8+4 \rho ^{12}-\rho ^{16}\right)\nonumber\\ &&+2 L \left(1-4 \rho ^4-2 \rho ^8-4 \rho ^{12}+\rho ^{16}\right)],
\\
\label{k}
k(r)&=&\frac{1}{16 h \rho ^4 \left(\rho ^4-1\right)^2}[\sqrt{2} g \left(1+4 \rho ^4+4 \rho ^{12}+\rho ^{16}\right)\nonumber\\ &&-2 L \left(1+4 \rho ^4-2 \rho ^8+4 \rho ^{12}+\rho ^{16}\right)],
\end{eqnarray}
where $r=\log{\rho}$ and $\kappa$ and $L$ satisfy,
\begin{equation}
\label{kappaL}
\kappa + L= \sqrt{2} g .
\end{equation}
In these expressions $g$ and $h$ are the gauge coupling of the vector fields\footnote{This constant enters in the superpotential even for vanishing profile for the vector fields.} and the topological mass of the 3-form potential, respectively, of minimal $\mathcal{N}=1$ 7d supergravity.

\subsection{The $r\to \infty$, AdS$_7$ limit}

When $r\rightarrow\infty$ the previous solution is asymptotically locally AdS$_7$, for any values of $\kappa$ and $L$ respecting the constraint given by their equation (4.27). The explicit way in which AdS$_7$ arises is as follows.

The $r\to \infty$ limit of the previous functions gives, for $g=2\sqrt{2} h$ \footnote{This value is fixed such that $X=1$ asymptotically.},
\begin{eqnarray}
\label{rbig}
X&\simeq& 1,\nonumber\\
e^{2U}&\simeq &\frac{\rho^4}{4}=\frac{e^{4r}}{4},\\
e^{2V}&\simeq &\frac{4}{h^2},\\
e^{2W}&\simeq &\frac{\rho^4}{4}=\frac{e^{4r}}{4},\\
k&\simeq &-\frac{\rho^4}{16}=-\frac{e^{4r}}{16},\\
l&\simeq &\frac{\rho^4}{16}=\frac{e^{4r}}{16}.
\end{eqnarray}
This gives for the 7d metric,
\begin{equation}
\label{metricAdS7DPrbig}
ds_7^2=\frac{e^{4r}}{L^2}ds^2(\text{AdS}_3)+\frac{4}{h^2}dr^2+\frac{e^{4r}}{\kappa^2}ds^2(\text{S}^3),
\end{equation}
in terms of unit radius S$^3$ and AdS$_3$ spaces. In turn, the 3-form potential is given by, 
\begin{equation}
B_3=\frac{\sqrt{2}g-2L}{16h}e^{4r} \Bigl(\text{vol}(\text{AdS}_3)-\text{vol}(\text{S}^3)\Bigr).
\end{equation}
For arbitrary $L$ and $\kappa$, the scalar curvature is 
\begin{equation}
R=-\frac32 e^{-4r}\Bigl(28 e^{4r}h^2+L^2-\kappa^2\Bigr),
\end{equation}
and thus asymptotes to that of an AdS$_7$ space of radius $1/h$. The geometry in the UV can thus be
{\it completed} by an AdS$_7$ space with vanishing 3-form potential, that solves the equations of motion and gives rise to an AdS$_7$ solution to massive IIA supergravity upon uplift to ten dimensions \cite{Passias:2015gya}.

\subsection{The $r\to 0$, AdS$_3\times$T$^4$ limit}

In turn, the $r\to 0$ limit of the expressions \eqref{X}-\eqref{k} is non-singular for the special value
\begin{equation}
\label{ele}
L=\frac{5g}{4\sqrt{2}},
\end{equation}
which is also the value for which the leading order behaviour of the scalar potential $\nu(X)$,
\begin{equation}
\nu (r)= \frac{h^{2/5}(5\sqrt{2}g-8L)^{8/5}}{2^{3/10} r^{16/5}}+\dots
\end{equation}
is non-singular. Note that from (\ref{kappaL}), 
\begin{equation}
\label{kappa}
\kappa=\frac{3g}{4\sqrt{2}}.
\end{equation}
Substituting these values in \eqref{X}-\eqref{k} and taking the $r\to 0$ limit, one finds
\begin{eqnarray}
\label{rsmall}
X&\simeq&\frac{2^{2/5}}{3^{1/5}},\nonumber\\
e^{2U}&\simeq &\frac{3^{2/5}}{2^{4/5}},\nonumber\\
e^{2V}&\simeq &\frac{2^8}{3g^2}\left(\frac{2}{3^3}\right)^{1/5},\nonumber\\
e^{2W}&\simeq &3^{2/5}2^{6/5}r^2.\nonumber\\
\end{eqnarray}
This gives, for the metric in \eqref{metricAdS7DP}  
\begin{equation}
\label{metricr0}
ds_7^2=\frac{2^{31/5}}{g^2}\Bigl(\frac{3^{2/5}}{5^2}ds^2(\text{AdS}_3)+\frac{4}{3^{8/5}}ds^2(\text{T}^4)\Bigr),
\end{equation}
and for the 3-form potential
\begin{equation}
B_3=-\frac12 \text{vol}(\text{AdS}_3)-4r^4 \text{vol}(\text{S}^3).
\end{equation}

As discussed in \cite{Dibitetto:2017tve}, this geometry is not a solution of 7d $\mathcal{N}=1$ minimal supergravity by itself, but rather the IR leading profile of the flow for $L$ and $\kappa$ given by (\ref{ele}), (\ref{kappa}). 


\subsection{Uplift to massive IIA}
\label{DPdetails-uplift}

7d $\mathcal{N}=1$ minimal supergravity can be consistently uplifted to massive IIA on a squashed S$^3$ \cite{Passias:2015gya}. The uplift formulae were provided in that reference. They read, in the parameterisation used in \cite{Cremonesi:2015bld} and for vanishing vector fields:
\begin{eqnarray}
\label{ds10a}
ds_{10}^2&=&\frac{16}{g}\pi\bigg(-\frac{\alpha}{\ddot{\alpha}}\bigg)^{1/2}X^{-1/2}ds^2_7\nonumber\\ &&+\frac{16}{g^3}\pi X^{5/2}\bigg[\bigg(-\frac{\ddot{\alpha}}{\alpha}\bigg)^{1/2}dz^2-\bigg(-\frac{\alpha}{\ddot{\alpha}}\bigg)^{1/2}\frac{\alpha\ddot{\alpha}}{\dot{\alpha}^2-2\alpha\ddot{\alpha}X^5}ds^2(\text{S}^2)\bigg]\\
e^{2\Phi}&=&\frac{X^{5/2}}{g^3}\frac{3^{8}2^{6}\pi^5}{\dot{\alpha}^2-2\alpha\ddot{\alpha}X^5}\bigg(-\frac{\alpha}{\ddot{\alpha}}\bigg)^{3/2} 
\\
B_2&=&\frac{2^3\sqrt{2}}{g^3}\Bigl(\frac{\pi\alpha\dot{\alpha}}{\dot{\alpha}^2-2\alpha\ddot{\alpha}X^5}-\pi z\Bigr)\text{vol}(\text{S}^2)\\
F_2&=&\Bigl(\frac{2^3\sqrt{2}}{g^3}F_0\frac{\pi\alpha\dot{\alpha}}{\dot{\alpha}^2-2\alpha\ddot{\alpha}X^5}+\frac{\ddot{\alpha}}{3^42\pi^2}\Bigr)\text{vol}(\text{S}^2)\\ \label{F410a}
F_4&=&\frac{2^3}{3^4\pi}\left[-\ddot{\alpha}dz\wedge B_{(3)}-\dot{\alpha}\text{d}B_{(3)}\right],
\end{eqnarray}
where in the last expression we have used the odd-dimensional self-duality condition \cite{Pilch:1984xy}
\begin{equation}
X^4\ast_7 \mathcal{F}_4=-2hB_{3}.
\end{equation}


\begin{thebibliography}{99}

\bibitem{Karch:2001cw}
  A.~Karch and L.~Randall,
  ``Localized gravity in string theory,''
  Phys.\ Rev.\ Lett.\  {\bf 87} (2001) 061601
  [hep-th/0105108].

\bibitem{Karch:2000gx}
  A.~Karch and L.~Randall,
  ``Open and closed string interpretation of SUSY CFT's on branes with boundaries,''
  JHEP {\bf 0106} (2001) 063
  [hep-th/0105132].

\bibitem{DeWolfe:2001pq}
  O.~DeWolfe, D.~Z.~Freedman and H.~Ooguri,
  ``Holography and defect conformal field theories,''
  Phys.\ Rev.\ D {\bf 66} (2002) 025009
  [hep-th/0111135].

\bibitem{DHoker:2007zhm}
  E.~D'Hoker, J.~Estes and M.~Gutperle,
  ``Exact half-BPS Type IIB interface solutions. I. Local solution and supersymmetric Janus,''
  JHEP {\bf 0706} (2007) 021
  [arXiv:0705.0022 [hep-th]].
  
\bibitem{DHoker:2007hhe}
  E.~D'Hoker, J.~Estes and M.~Gutperle,
  ``Exact half-BPS Type IIB interface solutions. II. Flux solutions and multi-Janus,''
  JHEP {\bf 0706} (2007) 022
  [arXiv:0705.0024 [hep-th]].

\bibitem{Dibitetto:2017tve}
  G.~Dibitetto and N.~Petri,
  ``BPS objects in D = 7 supergravity and their M-theory origin,''
  JHEP {\bf 1712} (2017) 041
  [arXiv:1707.06152 [hep-th]].
  
\bibitem{Dibitetto:2017klx}
  G.~Dibitetto and N.~Petri,
  ``6d surface defects from massive type IIA,''
  JHEP {\bf 1801} (2018) 039
  [arXiv:1707.06154 [hep-th]].
  
\bibitem{Dibitetto:2018iar}
  G.~Dibitetto and N.~Petri,
  ``Surface defects in the D4 $-$ D8 brane system,''
  JHEP {\bf 1901} (2019) 193
  [arXiv:1807.07768 [hep-th]].
  

\bibitem{Dibitetto:2018gtk}
  G.~Dibitetto and N.~Petri,
  ``AdS$_{2}$ solutions and their massive IIA origin,''
  JHEP {\bf 1905} (2019) 107
  [arXiv:1811.11572 [hep-th]].
  
\bibitem{Penin:2019jlf}
  J.~M.~Penin, A.~V.~Ramallo and D.~Rodriguez-Gomez,
  ``Supersymmetric probes in warped $AdS_6$,''
  arXiv:1906.07732 [hep-th].

  \bibitem{LMNR1}
  Y.~Lozano, N.~T.~Macpherson, C.~Nunez and A.~Ramirez,
  ``AdS$_3$ solutions in Massive IIA with small $\mathcal{N}=(4,0)$ supersymmetry,''
  arXiv:1908.09851 [hep-th].
  
  
\bibitem{Maldacena:1997de}
  J.~M.~Maldacena, A.~Strominger and E.~Witten,
  ``Black hole entropy in M theory,''
  JHEP {\bf 9712} (1997) 002
  [hep-th/9711053].

\bibitem{Vafa:1997gr}
  C.~Vafa,
  ``Black holes and Calabi-Yau threefolds,''
  Adv.\ Theor.\ Math.\ Phys.\  {\bf 2} (1998) 207
  [hep-th/9711067].
  
\bibitem{Minasian:1999qn}
  R.~Minasian, G.~W.~Moore and D.~Tsimpis,
  ``Calabi-Yau black holes and (0,4) sigma models,''
  Commun.\ Math.\ Phys.\  {\bf 209} (2000) 325
  [hep-th/9904217].
    
\bibitem{Castro:2008ne}
  A.~Castro, J.~L.~Davis, P.~Kraus and F.~Larsen,
  ``String Theory Effects on Five-Dimensional Black Hole Physics,''
  Int.\ J.\ Mod.\ Phys.\ A {\bf 23} (2008) 613
  [arXiv:0801.1863 [hep-th]].

\bibitem{Haghighat:2015ega}
  B.~Haghighat, S.~Murthy, C.~Vafa and S.~Vandoren,
  ``F-Theory, Spinning Black Holes and Multi-string Branches,''
  JHEP {\bf 1601} (2016) 009
  [arXiv:1509.00455 [hep-th]].

\bibitem{Couzens:2019wls}
  C.~Couzens, H.~h.~Lam, K.~Mayer and S.~Vandoren,
  ``Black Holes and (0,4) SCFTs from Type IIB on K3,''
  [arXiv:1904.05361 [hep-th]].

\bibitem{Kutasov:1998zh}
  D.~Kutasov, F.~Larsen and R.~G.~Leigh,
  ``String theory in magnetic monopole backgrounds,''
  Nucl.\ Phys.\ B {\bf 550} (1999) 183
  [hep-th/9812027].

\bibitem{Sugawara:1999qp}
  Y.~Sugawara,
  ``N = (0,4) quiver SCFT(2) and supergravity on AdS(3) x S**2,''
  JHEP {\bf 9906} (1999) 035
  [hep-th/9903120].

\bibitem{Larsen:1999dh}
  F.~Larsen and E.~J.~Martinec,
  ``Currents and moduli in the (4,0) theory,''
  JHEP {\bf 9911} (1999) 002
  [hep-th/9909088].

\bibitem{Okuyama:2005gq}
  K.~Okuyama,
  ``D1-D5 on ALE space,''
  JHEP {\bf 0512} (2005) 042
  [hep-th/0510195].

\bibitem{Douglas:1996uz}
  M.~R.~Douglas,
  ``Gauge fields and D-branes,''
  J.\ Geom.\ Phys.\  {\bf 28} (1998) 255
  [hep-th/9604198].
  
\bibitem{Haghighat:2013gba}
  B.~Haghighat, A.~Iqbal, C.~Kozcaz, G.~Lockhart and C.~Vafa,
  ``M-Strings,''
  Commun.\ Math.\ Phys.\  {\bf 334} (2015) no.2,  779
  [arXiv:1305.6322 [hep-th]].

\bibitem{Haghighat:2013tka}
  B.~Haghighat, C.~Kozcaz, G.~Lockhart and C.~Vafa,
  ``Orbifolds of M-strings,''
  Phys.\ Rev.\ D {\bf 89} (2014) no.4,  046003
  [arXiv:1310.1185 [hep-th]].
  
\bibitem{Kim:2015gha}
  J.~Kim, S.~Kim and K.~Lee,
  ``Little strings and T-duality,''
  JHEP {\bf 1602} (2016) 170
  [arXiv:1503.07277 [hep-th]].
\bibitem{Gadde:2015tra}
  A.~Gadde, B.~Haghighat, J.~Kim, S.~Kim, G.~Lockhart and C.~Vafa,
  ``6d String Chains,''
  JHEP {\bf 1802} (2018) 143
  [arXiv:1504.04614 [hep-th]].
  

\bibitem{Lawrie:2016axq}
  C.~Lawrie, S.~Schafer-Nameki and T.~Weigand,
  ``Chiral 2d theories from N = 4 SYM with varying coupling,''
  JHEP {\bf 1704} (2017) 111
  [arXiv:1612.05640 [hep-th]].
  
\bibitem{Couzens:2017way}
  C.~Couzens, C.~Lawrie, D.~Martelli, S.~Schafer-Nameki and J.~M.~Wong,
  ``F-theory and AdS$_{3}$/CFT$_{2}$,''
  JHEP {\bf 1708} (2017) 043
  [arXiv:1705.04679 [hep-th]].
  
\bibitem{Tong:2014yna}
  D.~Tong,
  ``The holographic dual of $AdS_{3} \times  S^{3} \times S^{3} \times S^{1}$,''
  JHEP {\bf 1404} (2014) 193
  [arXiv:1402.5135 [hep-th]].

\bibitem{Lozano:2015bra}
  Y.~Lozano, N.~T.~Macpherson, J.~Montero and E.~O~Colgain,
  ``New $AdS_3 \times S^2$ T-duals with $ \mathcal{N}=\left(0,4\right) $ supersymmetry,''
  JHEP {\bf 1508} (2015) 121
  [arXiv:1507.02659 [hep-th]].

\bibitem{Kelekci:2016uqv}
O.~Kelekci, Y.~Lozano, J.~Montero, E.~O~Colgain and M.~Park,
  ``Large superconformal near-horizons from M-theory,''
  Phys.\ Rev.\ D {\bf 93} (2016) no.8,  086010
  [arXiv:1602.02802 [hep-th]].
  
\bibitem{Hanany:2018hlz}
  A.~Hanany and T.~Okazaki,
  ``(0,4) brane box models,''
  JHEP {\bf 1903} (2019) 027
  [arXiv:1811.09117 [hep-th]].

\bibitem{Macpherson:2018mif}
  N.~T.~Macpherson,
  ``Type II solutions on AdS$_{3} \times$ S$^{3} \times$ S$^{3}$ with large superconformal symmetry,''
  JHEP {\bf 1905} (2019) 089
  [arXiv:1812.10172 [hep-th]].
  
\bibitem{LMNRPRL}
Y.~Lozano, N.~T.~Macpherson, C.~Nunez, A.~Ramirez, 
``1/4 BPS AdS$_3$/CFT$_2$,'' [arXiv:1909.09636 [hep-th]].  

\bibitem{LMNR2}
Y.~Lozano, N.~T.~Macpherson, C.~Nunez, A.~Ramirez, 
``Two dimensional  ${\mathcal N}=(0,4)$ quivers dual to $AdS_3$ solutions in massive IIA,'' [arXiv:1909.10510 [hep-th]].


\bibitem{Apruzzi:2013yva}
  F.~Apruzzi, M.~Fazzi, D.~Rosa and A.~Tomasiello,
  ``All $AdS_7$ solutions of type II supergravity,''
  JHEP {\bf 1404} (2014) 064
  [arXiv:1309.2949 [hep-th]].

\bibitem{Sfetsos:2010uq}
  K.~Sfetsos and D.~C.~Thompson,
  ``On non-abelian T-dual geometries with Ramond fluxes,''
  Nucl.\ Phys.\ B {\bf 846} (2011) 21
  [arXiv:1012.1320 [hep-th]].

\bibitem{Lozano:2016kum}
  Y.~Lozano and C.~Nunez,
  ``Field theory aspects of non-Abelian T-duality and $ \mathcal{N}  =$ 2 linear quivers,''
  JHEP {\bf 1605} (2016) 107
  [arXiv:1603.04440 [hep-th]].

\bibitem{Lozano:2016wrs}
  Y.~Lozano, N.~T.~Macpherson, J.~Montero and C.~Nunez,
  ``Three-dimensional $ \mathcal{N}=4 $ linear quivers and non-Abelian T-duals,''
  JHEP {\bf 1611} (2016) 133
  [arXiv:1609.09061 [hep-th]].

\bibitem{Lozano:2017ole}
  Y.~Lozano, C.~Nunez and S.~Zacarias,
  ``BMN Vacua, Superstars and Non-Abelian T-duality,''
  JHEP {\bf 1709} (2017) 000
  [arXiv:1703.00417 [hep-th]].

\bibitem{Itsios:2017cew}
  G.~Itsios, Y.~Lozano, J.~Montero and C.~Nunez,
  ``The AdS$_{5}$ non-Abelian T-dual of Klebanov-Witten as a $ \mathcal{N}=1 $ linear quiver from M5-branes,''
  JHEP {\bf 1709} (2017) 038
  [arXiv:1705.09661 [hep-th]].

\bibitem{Lozano:2018pcp}
  Y.~Lozano, N.~T.~Macpherson and J.~Montero,
  ``AdS$_{6}$ T-duals and type IIB AdS$_{6} \times$ S$^{2}$ geometries with 7-branes,''
  JHEP {\bf 1901} (2019) 116
  [arXiv:1810.08093 [hep-th]].

\bibitem{Putrov:2015jpa} 
  P.~Putrov, J.~Song and W.~Yan,
  ``(0,4) dualities,''
  JHEP {\bf 1603}, 185 (2016)
  [arXiv:1505.07110 [hep-th]].
  

\bibitem{Hanany:1997gh}
  A.~Hanany and A.~Zaffaroni,
  ``Branes and six-dimensional supersymmetric theories,''
  Nucl.\ Phys.\ B {\bf 529} (1998) 180
  [hep-th/9712145].

\bibitem{Brunner:1997gf}
  I.~Brunner and A.~Karch,
  ``Branes at orbifolds versus Hanany Witten in six-dimensions,''
  JHEP {\bf 9803} (1998) 003
  [hep-th/9712143].
 
\bibitem{Apruzzi:2017nck}
  F.~Apruzzi and M.~Fazzi,
  ``AdS$_{7}$/CFT$_{6}$ with orientifolds,''
  JHEP {\bf 1801} (2018) 124
  [arXiv:1712.03235 [hep-th]].

\bibitem{Apruzzi:2015wna}
  F.~Apruzzi, M.~Fazzi, A.~Passias, A.~Rota and A.~Tomasiello,
  ``Six-Dimensional Superconformal Theories and their Compactifications from Type IIA Supergravity,''
  Phys.\ Rev.\ Lett.\  {\bf 115} (2015) no.6,  061601
  [arXiv:1502.06616 [hep-th]].



\bibitem{Apruzzi:2015zna}
  F.~Apruzzi, M.~Fazzi, A.~Passias and A.~Tomasiello,
  ``Supersymmetric AdS$_{5}$ solutions of massive IIA supergravity,''
  JHEP {\bf 1506} (2015) 195
  [arXiv:1502.06620 [hep-th]].

\bibitem{Rota:2015aoa}
  A.~Rota and A.~Tomasiello,
  ``AdS$_{4}$ compactifications of AdS$_{7}$ solutions in type II supergravity,''
  JHEP {\bf 1507} (2015) 076
  [arXiv:1502.06622 [hep-th]].

\bibitem{Cremonesi:2015bld}
  S.~Cremonesi and A.~Tomasiello,
  ``6d holographic anomaly match as a continuum limit,''
  JHEP {\bf 1605} (2016) 031
  [arXiv:1512.02225 [hep-th]].

\bibitem{Bobev:2016phc}
  N.~Bobev, G.~Dibitetto, F.~F.~Gautason and B.~Truijen,
  ``Holography, Brane Intersections and Six-dimensional SCFTs,''
  JHEP {\bf 1702} (2017) 116
  [arXiv:1612.06324 [hep-th]].

\bibitem{Macpherson:2016xwk}
  N.~T.~Macpherson and A.~Tomasiello,
  ``Minimal flux Minkowski classification,''
  JHEP {\bf 1709} (2017) 126
  [arXiv:1612.06885 [hep-th]].
  
\bibitem{Gaiotto:2014lca}
  D.~Gaiotto and A.~Tomasiello,
  ``Holography for (1,0) theories in six dimensions,''
  JHEP {\bf 1412} (2014) 003
  [arXiv:1404.0711 [hep-th]].
  
 
\bibitem{Witten:1997yu}
  E.~Witten,
  ``On the conformal field theory of the Higgs branch,''
  JHEP {\bf 9707} (1997) 003
  [hep-th/9707093].

\bibitem{Diaconescu:1997gu}
  D.~E.~Diaconescu and N.~Seiberg,
  ``The Coulomb branch of (4,4) supersymmetric field theories in two-dimensions,''
  JHEP {\bf 9707} (1997) 001
  [hep-th/9707158].
	
\bibitem{Gaiotto:2009we}
  D.~Gaiotto,
  ``N=2 dualities,''
  JHEP {\bf 1208} (2012) 034
  doi:10.1007/JHEP08(2012)034
  [arXiv:0904.2715 [hep-th]].
	
\bibitem{Gaiotto:2009gz}
  D.~Gaiotto and J.~Maldacena,
 ``The Gravity duals of N=2 superconformal field theories,''
  JHEP {\bf 1210} (2012) 189
  doi:10.1007/JHEP10(2012)189
  [arXiv:0904.4466 [hep-th]].
	
\bibitem{Maldacena:2000mw}
  J.~M.~Maldacena and C.~Nunez,
  ``Supergravity description of field theories on curved manifolds and a no go theorem,''
  Int.\ J.\ Mod.\ Phys.\ A {\bf 16} (2001) 822
  doi:10.1142/S0217751X01003935, 10.1142/S0217751X01003937
  [hep-th/0007018].


\bibitem{Boonstra:1998yu}
  H.~J.~Boonstra, B.~Peeters and K.~Skenderis,
  ``Brane intersections, anti-de Sitter space-times and dual superconformal theories,''
  Nucl.\ Phys.\ B {\bf 533} (1998) 127
  [hep-th/9803231].
  
\bibitem{Brodie:1997wn}
  J.~H.~Brodie,
  ``Two-dimensional mirror symmetry from M theory,''
  Nucl.\ Phys.\ B {\bf 517} (1998) 36
  [hep-th/9709228].
  
\bibitem{Ito:1997ep}
  K.~Ito and N.~Maru,
  ``Matrix string theory from brane configuration,''
  Phys.\ Lett.\ B {\bf 426} (1998) 43
  [hep-th/9710029].
  
\bibitem{Bobev:2017uzs}
  N.~Bobev and P.~M.~Crichigno,
  ``Universal RG Flows Across Dimensions and Holography,''
  JHEP {\bf 1712} (2017) 065
  [arXiv:1708.05052 [hep-th]].
  
\bibitem{Nunez:2018ags}
  C.~Nu\~nez, J.~M.~Penin, D.~Roychowdhury and J.~Van Gorsel,
  ``The non-Integrability of Strings in Massive Type IIA and their Holographic duals,''
  JHEP {\bf 1806} (2018) 078
  [arXiv:1802.04269 [hep-th]].
  
  

  
\bibitem{Passias:2015gya}
  A.~Passias, A.~Rota and A.~Tomasiello,
  ``Universal consistent truncation for 6d/7d gauge/gravity duals,''
  JHEP {\bf 1510} (2015) 187
  [arXiv:1506.05462 [hep-th]].
  
\bibitem{Benini:2013cda}
  F.~Benini and N.~Bobev,
  ``Two-dimensional SCFTs from wrapped branes and c-extremization,''
  JHEP {\bf 1306} (2013) 005
  [arXiv:1302.4451 [hep-th]].
  
\bibitem{Casero:2006pt} 
  R.~Casero, C.~Nunez and A.~Paredes,
  ``Towards the string dual of N=1 SQCD-like theories,''
  Phys.\ Rev.\ D {\bf 73}, 086005 (2006)
  [hep-th/0602027].
  C.~Nunez, A.~Paredes and A.~V.~Ramallo,
  ``Unquenched Flavor in the Gauge/Gravity Correspondence,''
  Adv.\ High Energy Phys.\  {\bf 2010}, 196714 (2010)
  [arXiv:1002.1088 [hep-th]].

\bibitem{Giveon:1998ns}
  A.~Giveon, D.~Kutasov and N.~Seiberg,
  ``Comments on string theory on AdS(3),''
  Adv.\ Theor.\ Math.\ Phys.\  {\bf 2} (1998) 733
  [hep-th/9806194].

\bibitem{Seiberg:1999xz}
  N.~Seiberg and E.~Witten,
  ``The D1 / D5 system and singular CFT,''
  JHEP {\bf 9904} (1999) 017
  [hep-th/9903224].

\bibitem{Aharony:1999dw}
  O.~Aharony and M.~Berkooz,
  ``IR dynamics of D = 2, N=(4,4) gauge theories and DLCQ of 'little string theories',''
  JHEP {\bf 9910} (1999) 030
  [hep-th/9909101].
  
\bibitem{Kim:2007hv}
  H.~Kim, K.~K.~Kim and N.~Kim,
  ``1/4-BPS M-theory bubbles with SO(3) x SO(4) symmetry,''
  JHEP {\bf 0708} (2007) 050
  [arXiv:0706.2042 [hep-th]].

\bibitem{Macpherson:2015tka}
  N.~T.~Macpherson, C.~Nunez, D.~C.~Thompson and S.~Zacarias,
  ``Holographic Flows in non-Abelian T-dual Geometries,''
  JHEP {\bf 1511} (2015) 212
  [arXiv:1509.04286 [hep-th]].


\bibitem{Alvarez:1993qi}
  E.~Alvarez, L.~Alvarez-Gaume, J.~L.~F.~Barbon and Y.~Lozano,
  ``Some global aspects of duality in string theory,''
  Nucl.\ Phys.\ B {\bf 415} (1994) 71
  [hep-th/9309039].
  
\bibitem{Brown:1986nw}
  J.~D.~Brown and M.~Henneaux,
  ``Central Charges in the Canonical Realization of Asymptotic Symmetries: An Example from Three-Dimensional Gravity,''
  Commun.\ Math.\ Phys.\  {\bf 104} (1986) 207.


\bibitem{David:2002wn}
  J.~R.~David, G.~Mandal and S.~R.~Wadia,
  ``Microscopic formulation of black holes in string theory,''
  Phys.\ Rept.\  {\bf 369} (2002) 549
  [hep-th/0203048].
  
\bibitem{Pilch:1984xy}
  K.~Pilch, P.~van Nieuwenhuizen and P.~K.~Townsend,
  ``Compactification of $d=11$ Supergravity on S(4) (Or 11 = 7 + 4, Too),''
  Nucl.\ Phys.\ B {\bf 242} (1984) 377.

    \end{thebibliography}
\end{document}